\documentclass[aip,pof,preprint]{revtex4-2}

\draft 

\usepackage{amssymb}
\usepackage{amsmath}
\usepackage{graphicx}
\usepackage{subcaption}
\usepackage[colorlinks=true,linkcolor=blue,urlcolor=black,bookmarksopen=true]{hyperref}
\usepackage{bookmark}

\begin{document}
\newcommand{\Nu}{\mathit{Nu}}
\newcommand{\Ra}{\mathit{Ra}}
\newcommand{\Pran}{\mathit{Pr}}
\newcommand{\Rey}{\mathit{Re}}

\title{A global similarity correction for the RANS modeling of natural convection in unstably stratified flows} 

\author{Da-Sol Joo}
\email[]{wnekthf3818@postech.ac.kr}
\affiliation{Department of Mechanical Engineering, Pohang University of Science and Technology, Pohang, Gyeongbuk 37673,~South Korea}

\date{\today}

\begin{abstract}
This study proposes a global similarity correction for Reynolds-averaged Navier–Stokes (RANS) modeling of buoyancy effects in unstably stratified flows. 
Conventional two-equation RANS models (e.g., the \(k\)-\(\varepsilon\) model) lack a clear criterion for incorporating unstable buoyancy effects in their scale-determining equations (e.g., \(\varepsilon\)-equation). 
To address this gap, a global correction function is introduced, derived from a generalized algebraic formulation that incorporates available potential energy as an additional parameter. 
This function reproduces a global similarity law commonly observed in natural convection flows—for instance, the correlation among the Nusselt, Rayleigh, and Prandtl numbers, which can be approximately expressed as a single power law over a wide parameter range.
A calibration method is proposed in which an approximate analytical solution for Rayleigh–Bénard convection is obtained via equilibrium analysis, confirming that the proposed model captures similarity relations not addressed by conventional one-point closures.
Numerical results show significantly improved agreement with experimental data, accurately reproducing Nusselt number dependencies over broad ranges of Rayleigh and Prandtl numbers in unstably stratified flows, such as Rayleigh–Bénard convection and two types of internally heated convection. The method remains fully compatible with standard RANS frameworks and reverts to traditional turbulence treatments in shear-driven flows where buoyant effects are negligible. By introducing only a single, simple, algebraic global function in the conventional \(\varepsilon\)-equation, this approach significantly enhances the accuracy and robustness of buoyancy-driven turbulence simulations.
\end{abstract}

\pacs{}

\maketitle 

\section{Introduction}\label{sec:1}

Buoyancy-driven turbulent flows are ubiquitous in nature, occurring in contexts ranging from indoor air circulation to large-scale oceanic and atmospheric processes. They also play a central role in numerous engineering applications, including room heating, nuclear power plant safety, electronics cooling, and solar energy systems.

A classic example of natural convection is Rayleigh–Bénard convection, where a fluid layer is bounded by an isothermally cooled top wall and an isothermally heated bottom wall, separated by a distance \(L\) with a temperature difference \(\Delta\). The Rayleigh (\(\Ra\)) and Prandtl (\(\Pran\)) numbers characterize the flow. Once \(\Ra\) exceeds the critical value \(\Ra_c = 1708\) by roughly an order of magnitude, turbulence develops \cite{busse1981transition,siggia1994high}. From an engineering perspective, the primary goal is to predict the Nusselt number (\(\Nu\)), a dimensionless measure of time-averaged heat transfer.

In practical engineering applications, Reynolds-averaged Navier–Stokes (RANS) models are typically used to estimate time-averaged turbulent heat transfer. Although large-eddy simulation and direct numerical simulation have become increasingly powerful in academic research—due to an ability to resolve the detailed fluid motions of turbulent eddies—they remain prohibitively expensive, as the computational cost scales sharply with \(\Ra\). Consequently, RANS approaches continue to be among the most widely used methods for engineering design and analysis.

Several review articles discuss buoyancy-driven flows in the context of RANS modeling, including those by Hanjalić \cite{hanjalic2002one}, Launder \cite{launder2005rans}, Durbin \cite{durbin2018some}, and Hanjalić and Launder \cite{hanjalic2021reassessment}. Comprehensive treatments of buoyancy effects in RANS models can also be found in notable books by Rodi \cite{rodi1980turbulence}, Burchard \cite{burchard2007applied}, Durbin \cite{durbin2011statistical}, and Hanjalić and Launder \cite{hanjalic2011modelling}.

To simplify the discussion, consider a closed domain bounded by two horizontal walls at different temperatures, with gravity acting downward. Depending on whether the temperature gradient is aligned with or opposed to gravity, two distinct flow regimes arise: (i) stably stratified flow with top heating and bottom cooling, in which buoyancy dampens turbulent kinetic energy, and (ii) unstably stratified flow with top cooling and bottom heating, where buoyancy contributes positively to turbulent kinetic energy production.

In the present work, two-equation RANS models commonly used in engineering applications, such as the $k$–\(\varepsilon\) or $k$–\(\omega\) models, are considered.
The focus is on the $k$–\(\varepsilon\) model, with a brief discussion of its fundamental principles followed by an explanation of how buoyancy effects are treated. The simplified \(\varepsilon\)-equation of the standard \(k\)-\(\varepsilon\) RANS equations \cite{jones1972prediction}, including buoyancy effects \cite{rodi1980turbulence}, is:
\[
\frac{D \varepsilon}{D t}
= C_{\varepsilon1}\,\frac{\varepsilon}{k} 
\left(-\overline{u_i u_j} \,\frac{\partial U_i}{\partial x_j}\right)
+ C_{\varepsilon g}\,\frac{\varepsilon}{k} 
\left(- g_i \beta \,\overline{\theta u_i}\right)
- C_{\varepsilon2}\,\frac{\varepsilon^2}{k}
+ \cdots
\]
Here, \(U_i\) and \(u_i\) denote the mean and fluctuating velocities from the Reynolds decomposition, respectively, and \(\theta\) represents the temperature fluctuation. The overbar \(\overline{(\cdot)}\) indicates Reynolds averaging, so \(\overline{u_i u_j}\) and \(\overline{\theta u_i}\) represent the Reynolds stress (\(\mathrm{m^2/s^2}\)) and the turbulent heat flux (\(\mathrm{K \cdot m/s}\)), respectively. Additional variables include the turbulent kinetic energy \(k\) (\(\mathrm{m^2/s^2}\)), the dissipation rate \(\varepsilon\) (\(\mathrm{m^2/s^3}\)), gravitational acceleration \(g_i\) (\(\mathrm{m/s^2}\)), and the thermal expansion coefficient \(\beta\) (\(\mathrm{K^{-1}}\)). 
On the right-hand side of the \(\varepsilon\)-equation, the first two terms inside the blankets represent the production of turbulent kinetic energy due to velocity gradients and buoyancy, respectively.

Although the \(k\)-equation closely resembles the exact transport equation for turbulent kinetic energy and is thus relatively straightforward to interpret, the \(\varepsilon\)-equation is derived empirically and is less transparent. Textbooks and review articles on two-equation models \cite{speziale1990analytical,pope2000turbulent,wilcox1998turbulence,durbin2011statistical} note that the \(k\)-\(\varepsilon\) model is empirical in nature, and that understanding the \(\varepsilon\)-equation often requires examining the analytic solutions it reproduces. For instance, in decaying turbulence, given an initial condition, the standard model predicts $k(t)=k_0\,(t/t_0)^{-n}$ and $\varepsilon(t)=\varepsilon_0\,(t/t_0)^{n+1}$ with $n=1/(C_{\varepsilon2}-1)$, as time \(t\) progresses.
In homogeneous shear flow, the model predicts \(\left(Sk/\varepsilon\right)=\left(C_{\varepsilon2}-1\right)/\left(C_{\varepsilon1}-1\right)\) and that $k(t)$ grows exponentially, where \(S\) is the constant shear rate. The standard model constants \(C_{\varepsilon1}=1.44\) and \(C_{\varepsilon2}=1.92\) were established by matching these analytic solutions.

In the \(\varepsilon\)-equation, buoyancy effects are typically modeled in the same manner as velocity-gradient production. As discussed by \citet{gibson1978ground,durbin2011statistical}, this approach reproduces the behavior of stably stratified shear flows, where the opposing effects of buoyancy (which dampens turbulent kinetic energy) and shear (which generates turbulent kinetic energy) can be characterized by the flux Richardson number (\(\textit{Ri}_f\)). Linear stability theory indicates that once \(\textit{Ri}_f\) exceeds 0.25, the flow is stabilized; correspondingly, RANS models can be designed to suppress turbulence for \(\textit{Ri}_f > 0.25\).

However, when buoyancy acts as a source of turbulence in unstably stratified flows, there is little consensus on how to incorporate buoyancy in the \(\varepsilon\)-equation\cite{durbin2011statistical,durbin2018some}. As noted in several books \cite{hanjalic2011modelling,burchard2007applied} and papers \cite{henkes1991natural,peng1999computation,hanjalic1993computation}, the coefficient \(C_{\varepsilon g}\), which represents the buoyant effects in the \(\varepsilon\)-equation, exhibits wide variability (ranging from 0 to 1.44) in the literature and lacks a generally accepted value.
A common choice\cite{choi2012turbulence,kenjerevs1995prediction,ince1989computation,craft1996recent,dehoux2017elliptic} sets \(C_{\varepsilon g}=C_{\varepsilon1}=1.44\), whereas Markatos and Pericleous \cite{markatos1984laminar} used \(C_{\varepsilon g}=0\), arguing that no clear physical rationale supports a buoyancy production term in the \(\varepsilon\)-equation. Hanjalić and Vasić \cite{hanjalic1993computation} demonstrated that if the buoyant production term is omitted by setting \(C_{\varepsilon g}=0\) in flows heated from below with near-zero mean velocity, the \(\varepsilon\)-equation becomes dominated by a sink term, decays rapidly to zero, and incorrectly predicts unbounded growth of turbulent kinetic energy. As a result, an empirically optimized value of \(C_{\varepsilon g}=0.8\) was adopted. Another widely cited approach, based on Rodi’s argument \cite{rodi1980turbulence}, sets \(C_{\varepsilon g}\approx 0\) for horizontally heated boundary layers and \(C_{\varepsilon g}\approx 1\) for vertically heated boundary layers \cite{peng1999computation,lazeroms2013explicit}. Henkes \cite{henkes1991natural} refined this criterion further by proposing \(C_{\varepsilon g}=\tanh(u/v)\), where \(u\) and \(v\) are velocity components perpendicular and parallel to gravity, respectively. However, as Hanjalić and Vasić \cite{hanjalic1993computation} noted, setting \(C_{\varepsilon g}\approx 0\) is problematic, and \(\tanh(u/v)\) does not satisfy Galilean invariance. In commercial and open-source CFD software, \(C_{\varepsilon g}\) is typically either \(0\) by default with a \(\tanh(u/v)\) option (ANSYS Fluent) \cite{fluent2013ansys}, or set to \(1\) by default (OpenFOAM) \cite{openfoam2025}.

A key reason for the confusion surrounding unstable buoyancy effects in the \(\varepsilon\)-equation is the lack of a well-defined benchmark problem. For instance, in the standard model, the coefficients \(C_{\varepsilon1}\) and \(C_{\varepsilon2}\) (not related to buoyancy) were calibrated using benchmark problems such as homogeneous shear flow and decaying homogeneous turbulence. In contrast, benchmark problems or analytic solutions for unstable buoyancy effects have rarely been proposed, leaving the modeling of buoyant production terms on a less secure footing.

The absence of a benchmark problem for unstable buoyancy flows is largely due to the fact that buoyancy-generated turbulence does not exhibit the same self-similarity found in homogeneous shear flow. In homogeneous shear flow \cite{pope2000turbulent}, the system is governed primarily by the turbulent kinetic energy \(k\), the dissipation rate \(\varepsilon\), and the shear rate \(S\). This yields a self-similarity expressed as \(\varepsilon/k \sim S\), independent of the eddy length scale, indicating that it is a local relation not connected with global geometrical information.
By contrast, evidence for a locally defined self-similarity law in unstable buoyant flows is scarce.
Natural convection turbulence is typically characterized by dimensionless numbers such as the Nusselt ($\Nu$), Rayleigh ($\Ra$), and Prandtl ($\Pran$) numbers, with similarity often expressed through power-law relationships of the form \(\Nu \sim \Ra^{n} \Pran^{m}\). Since these parameters rely on global temperature differences and a characteristic length scale for the entire domain, this similarity is inherently nonlocal. Indeed, phenomenological turbulence theories \cite{grossmann2000scaling,ahlers2009heat} explain the observed power-law relations in terms of large-scale coherent flows on the order of the domain size. As a result, conventional two-equation RANS models, which inherently assume local self-similarity, struggle to justify modeling buoyant generation of turbulent kinetic energy in the same manner as shear-based generation.

This study aims to incorporate the “global similarity of unstable buoyant turbulence” while making only minimal modifications to conventional \(k\)-\(\varepsilon\) models. Specifically, a single global correction function that multiplies \(C_{\varepsilon g}\) in the \(\varepsilon\)-equation is introduced. This approach extends and generalizes previous work by \citet{joo2024reynolds}. 
The previous work also aimed to capture \(\Nu\)–\(\Ra\)–\(\Pran\) correlations in unstably stratified flows, but it relied on additional problem-specific variables and was applicable only to Rayleigh–Bénard convection.
The present study generalizes the previous work by relying exclusively on available potential energy as an additional variable in the correction function. Available potential energy, first proposed by Lorenz \cite{lorenz1955available}, represents the portion of a system's potential energy due to density variations that can be converted into kinetic energy, and it has been extensively used to parameterize buoyancy-driven geophysical flows \cite{taylor2023submesoscale,gayen2022rotating,tailleux2013available,wunsch2004vertical,peltier2003mixing}. Since available potential energy is defined as a unique and strictly positive value for a given global temperature field, incorporating it into the \(k\)-\(\varepsilon\) model preserves the model’s completeness and enables its application to a wide range of problems.

Existing RANS-based investigations into buoyancy-driven flows appear to have a rather limited ability to reproduce the $\Nu$–$\Ra$–$\Pran$ correlations described earlier, and they have generally been validated only over relatively narrow ranges of $\Ra$ and $\Pran$. Some examples, summarized again from the previous study by \citet{joo2024reynolds}, are as follows: Vertically heated natural convection in enclosures has been examined by Henkes \cite{henkes1991natural} for \(\Ra=10^{14} - 10^{15}\) and \(\Pran=0.7 - 7\), Peng \cite{peng1999computation} for \(\Ra=10^{14} - 10^{15}\), Kenjerevs \cite{kenjerevs1995prediction} for \(\Ra=10^{8} - 10^{10}\), and Dehoux \cite{dehoux2017elliptic} for \(\Ra=10^{11}\). Vertically heated natural convection in channels has been studied by Dol \cite{dol1999dns} for \(\Ra=10^{5} - 10^{7}\), Shin \cite{shin2008elliptic} for \(\Ra=10^{6}\), and Dehoux \cite{dehoux2017elliptic} for \(\Ra=10^{5} - 10^{7}\). Kenjerevs \cite{kenjerevs1995prediction} also investigated cases where the inner cylinder is heated and the outer cylinder is cooled, covering \(\Ra=10^{4} - 10^{9}\). In studies that specify only the \(\Ra\) range, \(\Pran=0.7\) was applied.
In summary, existing models have difficulty producing consistent results over a wide range of $\Ra$ and $\Pran$. If these limitations are overcome and the predicted $\Nu$–$\Ra$–$\Pran$ correlations can be guaranteed over a broad parameter range, model reliability is expected to improve significantly.

From an engineering perspective, there is reason to believe that optimizing a model to reproduce the $\Nu$–$\Ra$–$\Pran$ similarity relation in one problem can approximately improve its accuracy in other natural convection problems. This inference arises from several observed facts. Experiments by \citet{niemela2000turbulent} indicate that turbulent Rayleigh–Bénard convection follows an approximate \(\Nu \sim \Ra^{0.31}\) scaling law for \(\Ra \le 10^{13}\), and this correlation is nearly independent of the aspect ratio when the aspect ratio is sufficiently large \cite{ahlers2009heat}. Similar scaling laws are observed even when boundary conditions or geometry are modified. For example, a square cavity with differently heated parallel walls aligns directly with an Rayleigh–Bénard convection problem when the temperature gradient and gravity act in the same direction. Rotating the domain by 90 degrees transforms the problem into vertically heated convection, producing an average Nusselt number (for \(\Ra \lesssim 10^{11}\)) that is consistently about 20\% lower than in Rayleigh–Bénard convection. Therefore, although the prefactor differs, the power-law exponent remains unchanged, and the scaling law \(\Nu \sim \Ra^{0.3}\) continues to hold over a broad parameter range \cite{bejan2013convection,guo2015effect,chand2022effect}.
Another example involves natural convection driven by internal heating and cooling walls. The modified Rayleigh number \(\Ra^{\prime}\), based on a volumetric heat source, is defined as an alternative to \(\Ra\). In turbulent natural convection with internal heating, experimental results \cite{kulacki1977steady} show that \(\Nu\) scales approximately as \(\Nu \sim (\Ra^{\prime})^{0.23}\) for \(\Ra^{\prime} < 10^{12}\). Similar power-law behavior has been reported over a wide range of parameters and geometries, including cylinders and hemispheres \cite{zhang2015natural}. Substituting the heat balance condition \(\Ra^{\prime} \sim \Ra\,\Nu\) and rearranging yields \(\Nu \sim (\Ra^{\prime})^{0.23} \sim \Ra^{0.3}\), which closely resembles other natural convection correlations.

The modeling approach of this study is to introduce a global correction function in front of the buoyancy-related model constant \(C_{\varepsilon g}\) in the \(\varepsilon\)-equation, thereby analytically ensuring a proper \(\Nu\)--\(\Ra\)--\(\Pran\) correlation for Rayleigh–Bénard convection. This approach is loosely justified by the aforementioned observations that an approximately consistent similarity relation exists across various natural convection problems.

The remainder of this paper is organized as follows.  
Section~\ref{sec:2} derives the nonlocal similarity correction function, which is incorporated into the buoyancy-related term of the \(\varepsilon\)-equation. An equilibrium analysis confirms the resulting approximate power-law behavior, \(\Nu \sim \Ra^{n}\Pran^{m}\).  
Section~\ref{sec:3} presents the simulation results and compares them with existing models. The new model is first calibrated for Rayleigh–Bénard convection over a wide range of \(\Ra\) and \(\Pran\), then tested on two internally heated convection problems: one with top-only cooling and one with both top and bottom cooling.  
This study demonstrates that incorporating a global correction function preserves the desired scaling over a broad parameter range, thus improving reliability across different natural convection problems.
Finally, Section~\ref{sec:4} provides concluding remarks.

\section{Global Similarity Correction Function in the \(\varepsilon\)-Equation} \label{sec:2}

\subsection{Target of Modeling} \label{sec:2.1}

This study introduces a global similarity correction function \(f_{\varepsilon g}\) as an additional term in the conventional \(\varepsilon\)-equation, expressed as follows:
\begin{equation}
	\frac{D \varepsilon}{D t}
	= C_{\varepsilon1}\,\frac{\varepsilon}{k} 
	\left(-\overline{u_i u_j} \,\frac{\partial U_i}{\partial x_j}\right)
	+ C_{\varepsilon g} f_{\varepsilon g} \,\frac{\varepsilon}{k} 
	\left(- g_i \beta \,\overline{\theta u_i}\right)
	- C_{\varepsilon2}\,\frac{\varepsilon^2}{k}
	+ \cdots
\end{equation}
The specific formulation of \(f_{\varepsilon g}\) will be presented in Section \ref{sec:2.2}.

The remaining model equations combine (i) the standard  \(k\)-\(\varepsilon\) model by Launder and Sharma \cite{launder1974application} and (ii) the reduced algebraic turbulent heat flux model (AFM), along with a temperature variance transport equation proposed by Kenjereš \cite{kenjerevs1995prediction}. Although the gradient diffusion hypothesis (GDH) is widely used in commercial and open-source software, it fails to capture situations—such as Rayleigh–Bénard convection—where the temperature gradient is nearly zero but turbulent heat transfer remains significant.

As noted by Hanjalić \cite{hanjalic2002one}, GDH is not suitable for unstably stratified flows. In these flows, the time-averaged velocity is zero, and the temperature gradient in most of the domain (beyond the thin thermal boundary layers) remains very small, yet turbulent heat flux persists and maintains a large, nearly constant value in the center of domain. Consequently, GDH-based models—which represent turbulent heat flux as proportional to the local temperature gradient—break down in such cases.

An alternative is to derive turbulence models from the exact transport equation for the turbulent heat flux vector, which explicitly includes buoyancy effects. By applying the local-equilibrium assumption, it is possible to algebraically simplify the material derivative term. This approach underpins the algebraic heat flux model by Gibson and Launder \cite{gibson1978ground}, as well as a simplified version by Kenjeres \cite{kenjerevs1995prediction}. These models include a term proportional to gravity and temperature variance, enabling them to handle situations in which the temperature gradient is small.

All Reynolds-averaged equations for incompressible flow in this study are given below. The momentum and temperature transport equations are:
\begin{equation}
	\dfrac{D U_i}{D t}
	= - \dfrac{\partial P}{\partial x_i} 
	- g_i \beta \left( T - T_0 \right)
	+ \dfrac{\partial }{\partial x_j} 
	\left( \nu \dfrac{\partial U_i}{\partial x_j} -\overline{u_i u_j} \right),
\end{equation}
\begin{equation}\label{eq:3}
	\dfrac{D T}{D t}
	+ U_i \dfrac{\partial T}{\partial x_i}
	=
	Q
	+ \dfrac{\partial }{\partial x_i} 
	\left( \alpha \dfrac{\partial T}{\partial x_i} - \overline{\theta u_i} \right),
\end{equation}
Where \(P\) is the mean pressure per unit density, \(T\) is the mean temperature from Reynolds decomposition, \(T_0\) is the reference temperature in the Boussinesq approximation, \(Q\) (K/s) is the rate of temperature increase due to a volumetric heat source, and \(\alpha\) is the thermal diffusivity. 
If a domain is enclosed, \(T_0\) does not affect the net buoyancy force because it is offset by the normal force at the lower walls. For periodic boundaries perpendicular to gravity, \(T_0\) is typically set to the overall average temperature, making the net buoyancy force zero. 
For simplicity, \(T_0 = 0\)  is applied in the rest of the paper.

Following Launder and Sharma \cite{launder1974application}, the transport equations for turbulent kinetic energy \(k\) and its dissipation rate $\varepsilon+\varepsilon_0$ are:
\begin{align}
	\dfrac{D k}{D t}
	=& -\overline{u_i u_j} \dfrac{\partial U_i}{\partial x_j} 
	- g_i \beta \overline{\theta u_i}
	- \left( \varepsilon + \varepsilon_0 \right) 
	+ \dfrac{\partial }{\partial x_i} 
	\left[ \left(\nu + \dfrac{\nu_t}{\sigma_k}\right) \dfrac{\partial k}{\partial x_i}  \right],
\end{align}
\begin{align}
	\dfrac{D \varepsilon}{D t}
	=& 
	C_{\varepsilon1}\dfrac{\varepsilon}{k} \left[-\overline{u_i u_j} \dfrac{\partial U_i}{\partial x_j} \right]
	+ C_{\varepsilon g} f_{\varepsilon g}  \dfrac{\varepsilon}{k} \left[- g_i \beta \overline{\theta u_i}\right]
	- C_{\varepsilon2} f_\varepsilon \dfrac{\varepsilon^{2}}{k}
	+ \dfrac{\partial }{\partial x_i} 
	\left[ \left(\nu + \dfrac{\nu_t}{\sigma_\varepsilon}\right) \dfrac{\partial \varepsilon}{\partial x_i}  \right], 
\end{align}
where
\begin{equation}
	\varepsilon_0 = 2\nu \left( \dfrac{\partial k^{1/2}}{\partial x_i} \dfrac{\partial k^{1/2}}{\partial x_i} \right).
\end{equation}
The new function \(f_{\varepsilon g}\) is a correction term based on global potential energy, and this study specifically focuses on \(f_{\varepsilon g}\) and \(C_{\varepsilon g}\).
The turbulent viscosity and the Reynolds stress are given by
\begin{equation}
	\nu_t = C_{\mu} f_{\mu} \dfrac{k^{2}}{\varepsilon}, \quad
	\overline{u_{i} u_{j}}
	=
	\dfrac{2}{3}k \delta_{ij} - \nu_{t} \left( \dfrac{\partial U_i}{\partial x_j} + \dfrac{\partial U_j}{\partial x_i} \right).
\end{equation}
The damping functions are:
\begin{equation}
	f_{\varepsilon} = 1.0 - 0.3 \exp(-\Rey_{t}^{2}), \quad
	f_{\mu} = \exp \left[ -3.4 / \left( 1+ \Rey_{t}/50 \right)^{2}\right], \quad
	\text{with} \quad \Rey_{t} = \dfrac{k^{2}}{\nu \varepsilon}.
\end{equation}
The standard coefficients are $C_{\varepsilon1} = 1.44$, $C_{\varepsilon2} = 1.92$,	$C_{\mu} = 0.09$, $\sigma_k = 1.0$, $\sigma_\varepsilon = 1.3$.

If the turbulent heat flux is modeled with the GDH, it is written as:
\begin{equation}\label{eq:9}
	\overline{\theta u_i} = - \dfrac{\nu_{t}}{\Pran_{t}} \dfrac{\partial T}{\partial x_i},
\end{equation}
where \(\Pran_t \approx 1\) is often used for passive scalar problems by the Reynolds analogy.

However, this study employs the AFM proposed by Kenjeres et al. \cite{kenjerevs1995prediction}:
\begin{equation}\label{eq:10}
	\overline{\theta u_i^{\prime}} = -C_{\theta}\,\dfrac{k}{\varepsilon+\varepsilon_0}
	\left( 
	\overline{u^{\prime}_{i} u^{\prime}_{j}} \dfrac{\partial T}{\partial x_j}
	+ \xi \,\overline{\theta u^{\prime}_{j}} \dfrac{\partial U_i}{\partial x_j}
	+ \eta\, \beta g_i \,\overline{\theta^{2}}
	\right).
\end{equation}
Here, \(\overline{\theta^2}\) is the temperature variance (i.e., the Reynolds average of the square of the temperature fluctuation), obtained from its own transport equation:
\begin{equation}
	\dfrac{D \overline{\theta^{2}}}{D t}
	= - 2 \,\overline{\theta u^{\prime}_{i}} \dfrac{\partial T}{\partial x_i}
	- \dfrac{1}{R} \dfrac{\overline{\theta^2}}{k}\left( \varepsilon + \varepsilon_0\right) 
	+ \dfrac{\partial }{\partial x_i} 
	\left[ \left(\alpha + \dfrac{\nu_t}{\sigma_\theta}\right) \dfrac{\partial \overline{\theta^{2}}}{\partial x_i}  \right].
\end{equation}
The ratio of thermal to mechanical turbulent time scales, $R$, is treated as a constant when modeling the dissipation rate of $\overline{\theta}$. The model constants from \citet{kenjerevs1995prediction} are $C_{\theta} = 0.15$, $\xi=0.6$, $\eta=0.6$, $R=0.75$, $\sigma_\theta=1.0$.

As described in Section \ref{sec:1}, the available potential energy (APE) per unit volume and per unit density is newly introduced as a modeling variable. The total potential energy of a fluid per unit volume and per unit density across the entire domain $\Omega$ is
\begin{equation}
	E_p = \frac{g}{\rho_0 V} \int_\Omega \rho (x_i) \, z \,dV,
\end{equation}
where \(\rho(x_i)\) is the fluid density at position \(x_i\), and \(z \equiv -x_i g_i/|g|\) is the vertical coordinate in the direction opposite to gravity. The symbol \(V\) represents the total volume of the entire domain.

Adiabatic rearrangement\cite{lorenz1955available} is a hypothetical process in which fluid elements are rearranged so that denser elements lie below lighter ones, resulting in the rearranged density \(\rho^*(z)\). The total background potential energy is then
\begin{equation}
	E_b = \frac{g}{\rho_0 V} \int_\Omega \rho^{\ast}(x_i)\, z \,dV.
\end{equation}
The available potential energy per unit volume and per unit density, \(E_a\), is defined as
\begin{equation}
	E_a \equiv E_p - E_b = \frac{g}{\rho_0 V} \int_\Omega \left( \rho - \rho^{\ast} \right)\, z \,dV.
\end{equation}

In this study, the density is assumed to be $\rho = \rho_0 \left( 1 - \beta T \right)$ according to the Boussinesq approximation, leading to
\begin{equation}
	E_a = \frac{g \beta}{V} \int_\Omega \left(T^{\ast} - T \right)\, z \,dV,
\end{equation}
where \(T^{\ast}\) is the rearranged temperature field after the adiabatic rearrangement process.
During simulations, the rearrangement process for \(E_a\) is computed using a sorting algorithm. \(E_a\) is calculated at every simulation time (or iteration) step with an overall cost of \(\mathcal{O}(N \log N)\) for \(N\) grid elements, which is significantly lower than the cost of performing matrix inversions for the transport equations.
A more detailed mathematical description of APE can be found in \citet{winters1995available}.

\subsection{Turbulent Modeling of $f_{\varepsilon g}$} \label{sec:2.2}

The new global correction function, \(f_{\varepsilon g}\), is modeled as:
\begin{equation}\label{eq:16}
	f_{\varepsilon g} = 
	\left( \dfrac{k}{E_a} \right)^{a}
	\left( \dfrac{\max(0, -g_i \beta \,\overline{\theta u_i} )}{\varepsilon} \right)^{b}
	\left( \dfrac{k^{2}}{\nu \varepsilon} \right)^{d}
	\left( \dfrac{\nu}{\alpha} \right)^{e},
\end{equation}
where \(a\), \(b\), \(d\), and \(e\) are modeling constants. 
An empirical set of model constants is \(C_{\varepsilon g}=0.8\), \(a=0.75\), \(b=0.5\), \(d=0.5\), and \(e=0.4\).
The numerator of the second term on the right-hand side represents the buoyant kinetic energy production (clipped to zero if negative). All quantities in the brackets are dimensionless. The derivation of \(f_{\varepsilon g}\) will be presented in the remainder of this section.

\subsubsection{Dimensional analysis and definition of \(\boldsymbol{f_{\varepsilon g}}\)}

To derive \(f_{\varepsilon g}\), all independent variables that can influence its form are first considered. It is assumed that \(f_{\varepsilon g}\) depends on seven quantities:
\begin{itemize}
	\item Turbulent variables: \(k\) \((\mathrm{{m}^2/{s}^2})\), \(\varepsilon\) \((\mathrm{{m}^2/{s}^3})\), \(\overline{\theta^2}\) \((\mathrm{{K}^2})\)
	\item Material properties: \(g\beta\) \((\mathrm{{m}/({K}\cdot {s}^2)})\), \(\nu\) \((\mathrm{{m}^2/{s}})\), \(\alpha\) \((\mathrm{{m}^2/{s}})\)
	\item Global variable: \(E_a\) \((\mathrm{{m}^2/{s^2}})\)
\end{itemize}
Here, \(g\) and \(\beta\) are combined into a single variable \(g\beta\), because gravity appears only in buoyancy terms in this study.
Only if additional physics directly affected by gravity—such as an interaction of massive particles with a fluid—were present would \(g\) be treated separately from \(\beta\).

Since \(f_{\varepsilon g}\) is dimensionless but depends on three physical dimensions (length $\mathrm{m}$, time $\mathrm{s}$, and temperature $\mathrm{K}$) and seven total variables, the original seven variables are reduced to four dimensionless variables. Consequently, this yields:
\begin{equation}\label{eq:17}
	f_{\varepsilon g} = f \left( 
	\dfrac{k}{E_a}, \,
	\dfrac{\max(0, -g_i \beta\, \overline{\theta u_i} )}{\varepsilon}, \,
	\dfrac{k^{2}}{\nu \varepsilon}, \,
	\dfrac{\nu}{\alpha}
	\right).
\end{equation}
The second term represents the ratio of buoyant production to dissipation of turbulent kinetic energy. The third term is the turbulent Reynolds number \(\Rey_t\). The fourth term, $\nu/\alpha$, is the Prandtl number.

\begin{figure}[h]
	\centering
	\begin{subfigure}{0.3\textwidth}
		\centering
		\includegraphics[scale=0.5]{"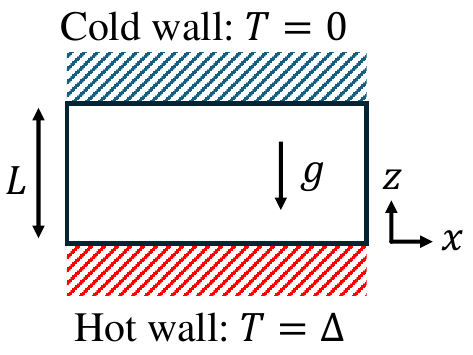"}
		\caption{}
	\end{subfigure}
	\begin{subfigure}{0.3\textwidth}
		\centering
		\includegraphics[scale=1.0]{"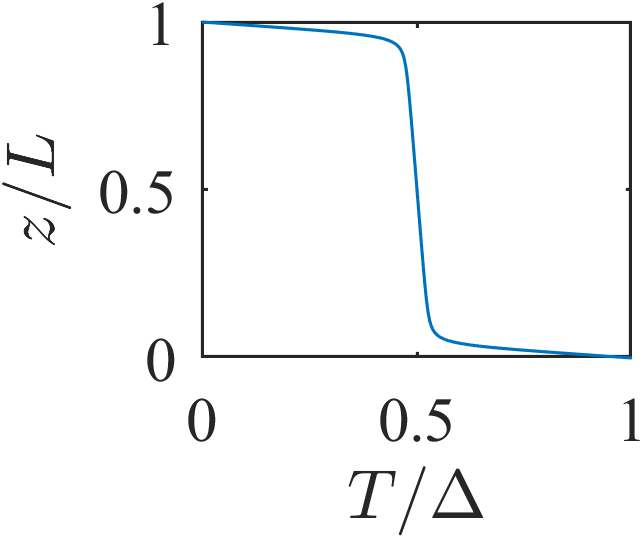"}
		\caption{}
	\end{subfigure}	
	\begin{subfigure}{0.3\textwidth}
		\centering
		\includegraphics[scale=1.0]{"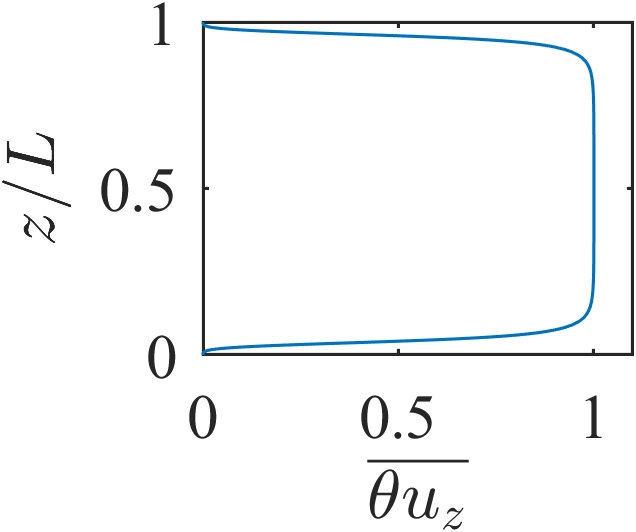"}
		\caption{}
	\end{subfigure}
	\caption{(a) Rayleigh–Bénard convection setup, (b) temperature distribution, and (c) turbulent heat flux (normalized by its maximum value).}
	\label{fig:1}
\end{figure}

As noted in Section~\ref{sec:1}, the function \(f_{\varepsilon g}\) is to be optimized for turbulent Rayleigh–Bénard convection. The goal is to confirm how the RANS model, when including \(f_{\varepsilon g}\), predicts the Nusselt number for given values of \(\Ra\) and \(\Pran\). Figure~\ref{fig:1}(a) shows the Rayleigh–Bénard setup with its boundary conditions. The dimensionless input parameters are the Rayleigh number (\(\Ra\)) and the Prandtl number (\(\Pran\)):
\begin{equation}
	\Ra = \dfrac{g\beta \,\Delta\, L^3}{\nu \,\alpha}, 
	\quad 
	\Pran = \dfrac{\nu}{\alpha}.
\end{equation}
Figures \ref{fig:1}(b) and \ref{fig:1}(c) show how the mean temperature and turbulent heat flux vary with height. The total heat flux \(q\), which is the sum of the conductive flux \(-\alpha\,(\partial T/\partial z)\) and the turbulent flux \(\overline{\theta u_z}\), remains constant along the vertical direction. The Nusselt number is then defined as
\begin{equation}
	\Nu = \dfrac{q\,L}{\alpha\,\Delta} 
	\quad \mathrm{for} \quad 
	q=-\alpha\frac{\partial T}{\partial z} + \overline{\theta u_z}.
\end{equation}
The thickness of the thermal boundary layer \(\delta_T\) satisfies \(\delta_T/L \sim \Nu^{-1}\), because \(\alpha(\Delta/\delta_T) \approx q\).

To simplify the problem, this study assumes all situations to be one-dimensional. In the Rayleigh–Bénard convection setup depicted in Figure~\ref{fig:1}(a), this means that all turbulent variables depend only on height \(z\).

\subsubsection{Bulk region simplification}
When \(\Nu\) is sufficiently large (\(\Nu \gg 1\)), the thermal boundary layer is thin \(\left(\delta_T/L \ll 1\right)\), and \(\overline{\theta u_z}\) can be assumed nearly constant in most of the domain (the “bulk” region), as shown in Figure~\ref{fig:1}(c). Under these conditions, the buoyant production \(-g_i \beta\, \overline{\theta u_i}\) is also nearly uniform in the bulk. Meanwhile, the total dissipation can be split between the bulk and the boundary layer, but the bulk contribution dominates \cite{grossmann2000scaling}. By volume-averaging the \(k\)-, \(\varepsilon\)-, and \(\overline{\theta^2}\)-equations across the entire domain and neglecting near-wall effects, the following is obtained:
\begin{equation}\label{eq:ode1}
	\dfrac{d k}{dt} = \left(\beta\,g\,\overline{\theta u_z}\right) - \varepsilon,
\end{equation}
\begin{equation}\label{eq:ode2}
	\dfrac{d \varepsilon}{dt} = 
	C_{\varepsilon g}\,f_{\varepsilon g} \,\left(\beta\,g\,\overline{\theta u_z}\right)\,\dfrac{\varepsilon}{k}
	- C_{\varepsilon 2}\,\dfrac{\varepsilon^2}{k},
\end{equation}
\begin{equation}\label{eq:ode3}
	\dfrac{d\overline{\theta^2}}{dt} = 
	2\,\overline{\theta u_z}\,\dfrac{\Delta}{L}
	- \dfrac{1}{R}\,\dfrac{\overline{\theta^2}}{k}\,\varepsilon,
\end{equation}
where
\begin{equation}
	\overline{\theta u_z} 
	= 
	\dfrac{2}{3}\,C_\theta \,\dfrac{k^2}{\varepsilon}
	\left(c' \,\dfrac{\Delta}{L}\right)
	+ C_{\theta}\,\eta\,\beta\,g\,\dfrac{k}{\varepsilon}\,\overline{\theta^2}.
\end{equation}
Here, \(k\), \(\varepsilon\), and \(\overline{\theta^2}\) are treated as spatially uniform, so they depend only on time. 
Consequently, the simplified model reduces to a set of ordinary differential equations (ODEs) in time. 
The parameter \(c'(\Delta/L)\) denotes the bulk temperature gradient (\(0 \le c' \le 1\)). 

In Rayleigh–Bénard convection, when the Rayleigh number is only slightly larger than its critical value, the fluid adopts laminar flow in the form of convection cells. As the Rayleigh number increases further, additional modes of large-scale convection cells can appear, resulting in a flow mixed in various directions rather than confined to a single direction\cite{busse1981transition}. Due to these characteristics, in high-Rayleigh-number turbulent Rayleigh–Bénard convection, the time-averaged velocity is zero, and thus the mechanical turbulence-production term arising from velocity gradients is also zero, as shown in the experimentally measured budget by \citet{togni2015physical}. Further details on this bulk-region simplification can be found in Joo et al.~\cite{joo2024reynolds}.

\subsubsection{Scaling analysis and inclusion of \(\boldsymbol{f_{\varepsilon g}}\)}
The next step is to verify the power-law behavior \(\Nu \sim \Ra^{n}\,\Pran^{m}\) when the new function \(f_{\varepsilon g}\) is added. Begin by applying a first-order Taylor expansion of \(\log(f_{\varepsilon g})\):
\begin{equation} \label{eq:fg1}
	\log\left(f_{\varepsilon g}\right) 
	= a\;\log \!\left(\dfrac{k}{E_a}\right) 
	+ b\;\log \!\left(\dfrac{g\,\beta\, \overline{\theta u_z}}{\varepsilon}\right) 
	+ d\;\log \!\left(\dfrac{k^{2}}{\nu\,\varepsilon}\right) 
	+ e\;\log \!\left(\dfrac{\nu}{\alpha}\right).
\end{equation}
By setting the time-derivative terms in the simplified ODEs to zero (i.e., at equilibrium), an algebraic solution for \(\Nu\) can be obtained, following the method described in Joo et al.~\cite{joo2024reynolds}.

Two cases are considered for estimating the available potential energy \(E_a\):
\begin{itemize}
	\item \textbf{Case 1:} If \(\partial T/\partial z \approx \Delta/L\) throughout the domain, then \(E_a \approx \tfrac{1}{6}\,g\,\beta\,\Delta\,L.\)
	\item \textbf{Case 2:} If \(\partial T/\partial z \approx 0\) in the bulk but \(\approx -\Delta/\delta_T\) in the boundary layer, then 
	\(\displaystyle E_a \sim g\,\beta\,\Delta\,\delta_T.\)  
	Since \(\alpha(\Delta/\delta_T)\approx \overline{\theta u_z}\), it follows that
	\(\displaystyle E_a \sim \frac{\alpha\,g\,\beta\,\Delta^2}{\,\overline{\theta u_z}\!}.\)
\end{itemize}
Focus is placed on the scaling relations for \(\Nu\) in terms of \(\Ra\) and \(\Pran\).

Substituting \(dk/dt = d\varepsilon/dt = d\overline{\theta^2}/dt = 0\) and \(f_{\varepsilon g}\) from Eq.~(\ref{eq:fg1}) into Eqs.~(\ref{eq:ode1}) through (\ref{eq:ode3}) produces a system of simultaneous equations (equilibrium equations). The solution of this system under equilibrium conditions represents the expected convergence outcome of the RANS model. 
This approach, referred to as “equilibrium analysis,” has been widely applied in RANS modeling to understand how the RANS solution responds to imposed forcing, as explained by \citet{durbin2011statistical}.

After moving the sink terms to the left-hand side and taking the logarithm of both sides, the system becomes entirely linear. Consequently, a unique set of solutions \((k, \varepsilon, \overline{\theta^2})\) emerges. Then, nondimensionalization of this solution \((k, \varepsilon, \overline{\theta^2})\) yields an analytical \(\Nu\)--\(\Ra\)--\(\Pran\) correlation.
The series of steps involved in deriving these equilibrium equations is also described in detail in the previous study by \citet{joo2024reynolds}.

Solving the equilibrium equations yields:
\begin{align}\label{eq:Equil 1}
	\Nu & \sim 
	C_{\varepsilon g} ^{-\left( \tfrac{1}{a+d} \right)}\,
	\Ra^{\left( \tfrac{a}{2a+2d} \right) }\,
	\Pran^{\left( \tfrac{a+2d-2e}{2a+2d} \right) } 
	&\text{for}&\quad
	E_a  \approx \frac{1}{6} g \beta \Delta L, \\
	\label{eq:Equil 2}
	\Nu & \sim 
	C_{\varepsilon g}^{-\left( \tfrac{1}{2a+d}\right) }\,
	\Ra^{\left( \tfrac{a}{4a+2d}\right) }\,
	\Pran^{\left( \tfrac{a+2d-2e}{4a+2d} \right) }
	&\text{for}&\quad
	E_a \sim g\beta \Delta \delta_T.
\end{align}
Because \(k\) and \(\overline{\theta^2}\) are linearly proportional once \(\overline{\theta u_z}\) is eliminated from the \(k\)- and \(\overline{\theta^2}\)-equations, the coefficient \(c'\) does not affect the power-law exponents. 
Also, the exponent \(b\) does not appear in the final scaling relations because the buoyant production is balanced by the dissipation rate.

\subsubsection{Role of the new global parameter}
In this study, the global variable \(E_a\) is introduced into the model for the first time, and the constant \(a\) governs how \(f_{\varepsilon g}\) depends on \(E_a\). From Eqs.~\eqref{eq:Equil 1} and \eqref{eq:Equil 2}, it is evident that if \(a=0\), the Rayleigh-number exponent becomes zero. When \(a=0\), the “equilibrium solution” becomes a line passing through the point \(k=\varepsilon=\overline{\theta^{2}}=0\), rather than a unique non-zero point. In other words, without a global parameter, the simplified ODE system cannot predict a unique \(\Nu\) for a given \(\Ra\) and \(\Pran\). This limitation was also noted in the earlier work by \citet{joo2024reynolds}, which provides a more detailed discussion.

In summary, the new model function \(f_{\varepsilon g}\) in Eq.~(\ref{eq:16}) is proposed in an exponential form to replicate the similarity relation \(\Nu \sim \Ra^{n}\Pran^{m}\) in Rayleigh–Bénard convection by incorporating the global parameter \(E_a\).

\subsection{Constraints for $f_{\varepsilon g}$} \label{sec:2.3}

This section proposes constraints on \(f_{\varepsilon g}\) to determine its model constants \(a\), \(b\), \(d\), and \(e\).

\subsubsection{Model behavior as gravity approaches zero}
The \(\varepsilon\)-equation that includes \(f_{\varepsilon g}\) can be written as
\begin{align}
	\dfrac{D\varepsilon}{Dt}=
	\underset{\displaystyle{ \sim \left( g \beta \right) ^{-a + b + 1}}}
	{\underbrace{
			C_{\varepsilon g} 
			\left[ 
			\left( \dfrac{k}{E_a} \right)^{a}
			\left( \dfrac{\max(0, -g_i \beta \,\overline{\theta u_i} )}{\varepsilon} \right)^{b}
			\left( \dfrac{k^{2}}{\nu \varepsilon} \right)^{d}
			\left( \dfrac{\nu}{\alpha} \right)^{e} 
			\right] 
			\dfrac{\varepsilon}{k} \,\left[- g_i \beta \,\overline{\theta u_i}\right]
	}}
	+ \cdots
\end{align}
Here, \(E_a\) is proportional to \((g\beta)\). 

As \(g \to 0\), the buoyant term must also vanish. This requirement leads to the condition
\begin{equation}
	-a + b + 1 > 0.
\end{equation}

\subsubsection{Monotonic increase of \(\Nu\) with \(\Ra\)}
From Eqs.~(\ref{eq:Equil 1}) and (\ref{eq:Equil 2}), the dependence of \(\Nu\) on \(\Ra\) is estimated as
\begin{equation}
	\Nu \;\sim\; \Ra^{ {\left( \dfrac{a}{2a+2d}\right) } }
	\quad\text{or}\quad 
	\Ra^{\left( \dfrac{a}{4a+2d}\right) }.
\end{equation}
These exponents must be greater than zero to satisfy the laws of thermodynamics. Under this criterion, however, the constant \(a\) can still be either positive or negative.

\subsubsection{Convergence criterion}

In Section \ref{sec:2.2}, the model system is simplified to a set of three ODEs. The objective of the modeling is for this dynamic system to always converge to the designed equilibrium solution, regardless of the initial values of \(k\), \(\varepsilon\), and \(\overline{\theta^2}\). To satisfy this convergence requirement and simultaneously meet the “monotonic increase of \(\Nu\) with \(\Ra\)” condition from the previous section, the model constant \(a\) in \(f_g\) must satisfy the following constraint:
\begin{equation}
	a > 0.
\end{equation}
The remainder of this section will derive this condition.

This stability condition is derived as follows: in the simplified model system given by Eqs.~(\ref{eq:ode1})--(\ref{eq:ode3}), the equilibrium point must be a ``stable node'', meaning that all real parts of the eigenvalues of the Jacobian matrix are negative. In other words, small deviations from the equilibrium solution should converge back to it. With this criterion, convergence near the equilibrium solution is ensured; however, it is a necessary condition rather than a sufficient one. This criterion was previously proposed and described in detail by \citet{joo2024reynolds}.

Note that analyzing the model’s stability using time-dependent ODEs is different from the so-called unsteady RANS approach. This study aims to develop a RANS model that predicts the same stable solution whether it is solved transiently (by including a time-derivative term) or in steady-state form (with time derivatives set to zero). Clearly, if a transient RANS simulation yields a stable solution where all time derivatives—including those for velocity, temperature, and other turbulent variables—become zero, that solution also satisfies the steady-state equations. If the stable node criterion for the time-dependent ODEs is met, solving the equations in steady-state form will, at least approximately, ensure model convergence. A detailed proof is provided in the following section.

First, an analysis is conducted to identify the factors that influence the convergence of the model. In short, the stability of the simplified model is independent of \(\Ra\), \(\Pran\), and the model constant \(C_{\varepsilon g}\). Instead, stability depends solely on the constants \(a\), \(b\), and \(d\) in \(f_{\varepsilon g}\). To verify this, the ODEs in Eqs.~(\ref{eq:ode1})--(\ref{eq:ode3}) are converted into dimensionless form.

If the equilibrium solution of the simplified ODEs is given by
\begin{equation}
	\Nu \sim C_{\varepsilon g}^{\gamma} \,\Ra^{n}\,\Pran^{m},
\end{equation}
where
\begin{equation}\label{eq:ODE exponents}
	\left\lbrace 
	\begin{matrix}
		\gamma=-\dfrac{1}{\,a+d\,}, \quad n=\dfrac{a}{2a+2d}, \quad m=\dfrac{a+2d-2e}{2a+2d} 
		& \mathrm{for} & E_a\approx \dfrac{1}{6}\, g\, \beta\, \Delta\, L, \\[6pt]
		\gamma=-\dfrac{1}{\,2a+d\,}, \quad n=\dfrac{a}{4a+2d}, \quad m=\dfrac{a+2d-2e}{4a+2d}
		& \mathrm{for} & E_a\approx \dfrac{\alpha\,g\,\beta\,\Delta^2}{\overline{\theta u_z}},
	\end{matrix}
	\right.
\end{equation}
then the following variable transformations are defined:
\begin{equation}
	k = \dfrac{\nu^2}{L^2} \;C_{\varepsilon g}^{\gamma} \;\Ra^{\,n+1} \;\Pran^{\,m-1} \,\phi_1,
\end{equation}
\begin{equation}
	\varepsilon = \dfrac{\nu^3}{L^4} \;C_{\varepsilon g}^{\gamma} \;\Ra^{\,n+1.5} \;\Pran^{\,m-1.5} \,\phi_2,
\end{equation}
\begin{equation}
	\overline{\theta^2} = \Delta^2 \;C_{\varepsilon g}^{\gamma} \;\Ra^{\,n} \;\Pran^{\,m} \,\phi_3,
\end{equation}
\begin{equation}
	t= \dfrac{L^2}{\nu} \;\Ra^{-0.5} \,\Pran^{\,0.5} \;\tau.
\end{equation}
Under these transformations, the ODEs become
\begin{equation}\label{eq:ode nond1}
	\dfrac{d \phi_1}{d \tau} =
	\left[  \dfrac{2}{3}\,C_{\theta}\,c^{\prime}\,\dfrac{\phi_1^2}{\phi_2} 
	+ C_{\theta}\,\eta\,\dfrac{\phi_1\,\phi_3}{\phi_2} \right]  - \phi_2,
\end{equation}
\begin{equation}\label{eq:ode nond2}
	\dfrac{d \phi_2}{d \tau} = 
	f_{\varepsilon g}^{\ast} \left[  
	\dfrac{2}{3}\,C_{\theta}\,c^{\prime}\,\dfrac{\phi_1^2}{\phi_2} 
	+ C_{\theta}\,\eta\,\dfrac{\phi_1\,\phi_3}{\phi_2} 
	\right] 
	\dfrac{\phi_2}{\phi_1}  
	- C_{\varepsilon 2}\,\dfrac{\phi_2^2}{\phi_1},
\end{equation}
\begin{equation}\label{eq:ode nond3}
	\dfrac{d \phi_3}{d \tau} =
	2\,\left[  
	\dfrac{2}{3}\,C_{\theta}\,c^{\prime}\,\dfrac{\phi_1^2}{\phi_2} 
	+ C_{\theta}\,\eta\,\dfrac{\phi_1\,\phi_3}{\phi_2} 
	\right]  
	- \dfrac{1}{R}\,\dfrac{\phi_2\,\phi_3}{\phi_1},
\end{equation}
where
\begin{equation}\label{eq:ode nond4}
	f_{\varepsilon g}^{\ast} =
	\left\lbrace 
	\begin{matrix}
		\left(6 \,\phi_1\right) ^{a} 
		\left( 
		\dfrac{2}{3}\,C_{\theta}\,c^{\prime}\,\dfrac{\phi_1^2}{\phi_2^2} 
		+ C_{\theta}\,\eta\,\dfrac{\phi_1\,\phi_3}{\phi_2^2} 
		\right)^b 
		\left(\dfrac{\phi_1^{2}}{\phi_2}\right)^{d} 
		& \mathrm{for} & E_a\approx \dfrac{1}{6}\, g\, \beta\, \Delta\, L, \\[12pt]
		\left(\phi_1\right) ^{a} 
		\left(
		\dfrac{2}{3}\,C_{\theta}\,c^{\prime}\,\dfrac{\phi_1^2}{\phi_2^2} 
		+ C_{\theta}\,\eta\,\dfrac{\phi_1\,\phi_3}{\phi_2^2} 
		\right)^{(a+b)} 
		\left(\dfrac{\phi_1^{2}}{\phi_2}\right)^{d}  
		& \mathrm{for} & E_a\approx \dfrac{\alpha\,g\,\beta\,\Delta^2}{\overline{\theta u_z}},
	\end{matrix}
	\right.
\end{equation}
Notably, the constant \(C_{\varepsilon g}\) and the input parameters \(\Ra\) and \(\Pran\) disappear in Eqs.~(\ref{eq:ode nond1})--(\ref{eq:ode nond4}), implying that the stability of the model does not depend on them. If the constants \(C_{\varepsilon 2}\), \(C_{\theta}\), \(\eta\), and \(R\) from earlier studies are considered fixed, whether the equilibrium point is a stable node depends only on the model constants \(a\), \(b\), and \(d\).

\begin{figure}[h]
	\centering
	\includegraphics[scale=1.0]{"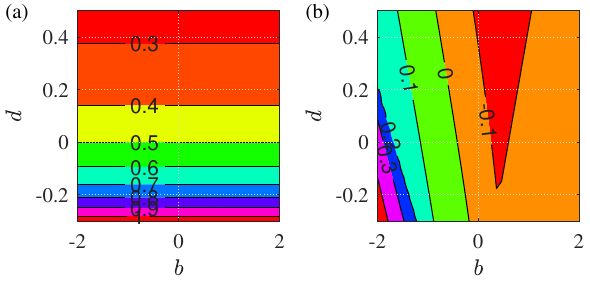"}
	\caption{For \(a = 0.75\) fixed, the contour plots as functions of \(b\) and \(d\) illustrate: (a) the exponent \(n\) in \(\Nu \sim \Ra^n\) from Eq.~(\ref{eq:ODE exponents}); and (b) the maximum real part of the eigenvalues at the equilibrium point.}
	\label{fig:2}
\end{figure}
\begin{figure}[h]
	\centering
	\includegraphics[scale=1.0]{"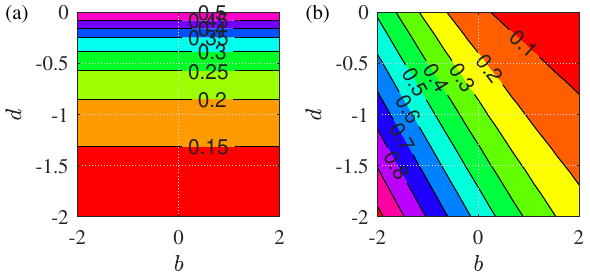"}
	\caption{For \(a = -0.75\) fixed, the contour plots as functions of \(b\) and \(d\) illustrate: (a) the exponent \(n\) in \(\Nu \sim \Ra^n\) from Eq.~(\ref{eq:ODE exponents}); and (b) the maximum real part of the eigenvalues at the equilibrium point.}
	\label{fig:3}
\end{figure}

Figure~\ref{fig:2}(a) shows the exponent \(n=a/(4a+2d)\)  for \(\Nu \sim \Ra^{n}\) versus \(b\) and \(d\), with \(a = 0.75\) fixed. Here, \(c^{\prime} = 0\) and \(E_a = g \beta \Delta \delta_T\) are assumed. The monotonic-increase condition \(n > 0\) is satisfied in the indicated range of \(b\) and \(d\). Within that same range, Figure~\ref{fig:2}(b) shows the maximum real part of the equilibrium point’s eigenvalues. For the convergence criterion to be met, this maximum real part must be negative.

On the other hand, Figure~\ref{fig:3} shows (a) the \(\Ra\) exponent and (b) the maximum real part of the eigenvalues as functions of \(b\) and \(d\) for \(a = -0.75\). When \(a < 0\), there is no suitable combination of \(b\) and \(d\) that satisfies both the monotonic-increase condition and the stable-node criterion.

For this reason, the model constant \(a\) in \(f_g\) must be greater than zero.
If the model constants \(a\), \(b\), and \(d\) are too large, the exponential terms in $f_{\varepsilon g}$ will change significantly at each simulation step, causing numerical instability. Conversely, if \(a\), \(b\), and \(d\) are too small, the calculation converges extremely slowly. An empirically determined set of model constants is \(C_{\varepsilon g} = 0.8\), \(a = 0.75\), \(b = 0.5\), \(d = 0.5\), and \(e = 0.4\).

\subsubsection{Note on Steady-State Simulations} \label{sec:2.3.4}

In the previous section on the convergence criterion, the model was approximated by spatially averaged ODEs across the entire domain, including \(d/dt\) terms for \(k\), \(\varepsilon\), and \(\overline{\theta^2}\), to analyze the model’s inherent stability. At equilibrium, \(dk/dt = d\varepsilon/dt = d\overline{\theta^2}/dt = 0\). This property implies that, with a suitable time-discretization scheme, a transient simulation incorporating these time derivatives will converge to a unique stable solution.

The discussion now shifts to the convergence behavior in steady-state simulations, where all the time derivative terms of \(k\), \(\varepsilon\), and \(\overline{\theta^2}\) are set to zero (\(d/dt = 0\)). Because the transport equations for \(k\) and \(\varepsilon\) contain these variables in both numerators and denominators, an iterative method that progressively projects the values toward the solution at each step is required. Let \(k_{n}\), \(\varepsilon_{n}\), and \(\overline{\theta^2}_{n}\) denote the turbulence variables at step \(n\), and \((n+1)\) represent the next iteration.

In this study, the \(k\)-equation is implemented as:
\begin{equation}
	\dfrac{k_{n+1}-k_{n}}{\Delta t} + U_{i}^{(n)}\dfrac{\partial k_{n+1}}{\partial x_i} 
	= \left( P + G \right) _{n}
	- (\varepsilon+\varepsilon_0)_{n}\,\dfrac{k_{n+1}}{k_{n}} 
	+ \nabla^{2} k_{n+1}.
\end{equation}
Here, \(P\equiv-\overline{u_i u_j}(\partial U_i / \partial x_j)\) is the mechanical production of turbulent kinetic energy, and \(G\equiv-g_i \beta \overline{\theta u_i}\) is the buoyancy-induced production, both evaluated at the previous step \(n\). The time-derivative term is included to illustrate the implicit Euler time-discretization used in this study. In a steady-state simulation, however, that time-derivative term is set to zero.

Similarly, the remaining transport equations can be written as:
\begin{equation}
	\dfrac{\varepsilon_{n+1}-\varepsilon_{n}}{\Delta t} 
	+ U_{i}^{(n)}\dfrac{\partial \varepsilon_{n+1}}{\partial x_i}  
	= \left(C_{\varepsilon 1}\,P + C_{\varepsilon g}\,f_{\varepsilon g}\,G\right)_{n}\,\dfrac{\varepsilon_{n}}{k_{n}}
	- C_{\varepsilon 2}\,\dfrac{\varepsilon_{n}}{k_{n}} \,\varepsilon_{n+1}
	+ \nabla^{2} \varepsilon_{n+1},
\end{equation}
\begin{equation}
	\dfrac{\overline{\theta^2}_{n+1}-\overline{\theta^2}_{n}}{\Delta t} 
	+ U_{i}^{(n)}\dfrac{\partial \overline{\theta^2}_{n+1}}{\partial x_i} 
	=  \Theta_{n} 
	- \dfrac{1}{R}\,\dfrac{(\varepsilon+\varepsilon_0)_{n}}{k_{n}}\,\overline{\theta^2}_{n+1}
	+ \nabla^{2} \overline{\theta^2}_{n+1},
\end{equation}
where \(\Theta\equiv-2\overline{\theta u_i}(\partial T /\partial x_i)\) represents the production of temperature variance.

Because all three equations can be discretized into linear forms in terms of \(k_{n+1}\), \(\varepsilon_{n+1}\), and \(\overline{\theta^2}_{n+1}\), the resulting linear system can be solved by matrix inversion at each iteration.

The key question is whether this iterative procedure can converge in a manner consistent with the transient approach. To analyze this in a simplified way, the equations can be reduced by omitting convection and diffusion terms. For instance, the turbulent kinetic energy equation might be written:
\begin{equation}
	\dfrac{\Delta k}{\Delta t} = f_{k} - g_{k}\,\left(k_{n} + \Delta k\right),
\end{equation}
where
\begin{equation}
	f_{k} = \left( P + G\right)_{n}, 
	\quad 
	g_{k} = \dfrac{(\varepsilon+\varepsilon_0)_{n}}{k_{n}}, 
	\quad 
	\Delta k = k_{n+1} - k_{n}.
\end{equation}
Here, \(f_{k}\) denotes terms that do not depend on \(k_{n+1}\), while \(g_{k}\) represents coefficients multiplying \(k_{n+1}\). 

In the steady-state iteration process, the following relation holds:
\begin{equation}
	0=f_{k} - g_{k}\,\left(k_{n} + \Delta k\right),
\end{equation}
so that the increment \(\Delta k\) is given by
\begin{equation}\label{eq:47}
	\Delta k 
	= 
	\dfrac{f_{k}}{g_{k}} - k_{n}.
\end{equation}
Analogous expressions can be written for \(\varepsilon\) and \(\overline{\theta^2}\):
\begin{equation}\label{eq:48}
	\Delta \varepsilon =  \dfrac{f_{\varepsilon}}{g_{\varepsilon}} - \varepsilon_{n},
	\quad
	\Delta \overline{\theta^2} =  \dfrac{f_{\theta}}{g_{\theta}} - \overline{\theta^2}_{n}.
\end{equation}
where
\begin{equation}
	\begin{matrix}
		f_\varepsilon = \left( C_{\varepsilon 1}P + C_{\varepsilon g}f_{\varepsilon g}G\right)_{n}, & 
		g_\varepsilon = C_{\varepsilon 2} \dfrac{\varepsilon_{n}}{k_{n}}, & 
		\Delta \varepsilon = \varepsilon_{n+1} - \varepsilon_{n}, \\[4pt]
		f_\theta = \Theta_{n}, & 
		g_\theta = \dfrac{1}{R} \dfrac{(\varepsilon+\varepsilon_0)_{n} }{k_{n}}, & 
		\Delta \overline{\theta^2} = \overline{\theta^2}_{n+1} - \overline{\theta^2}_{n}. 
	\end{matrix}
\end{equation}

Iterations proceed by updating these increments until convergence. Near the convergence point, stability requires (in a simple example) that if \(k_{n}\) is larger than the equilibrium value, then \(\Delta k\) should be negative. This idea generalizes to the full three-equation system and implies that the Jacobian matrix of this steady-state iteration must form a stable node. The Jacobian matrix for the steady-state system $J_S$ from Eqs.~(\ref{eq:47}) and (\ref{eq:48}) is given by:
\begin{equation}
	J_{S} = 
	\begin{bmatrix}
		\dfrac{f_{k,k}}{g_k} - \dfrac{f_{k}\,g_{k,k}}{g_k^{2}} - 1 
		&  \dfrac{f_{k,\varepsilon}}{g_k} - \dfrac{f_{k}\,g_{k,\varepsilon}}{g_k^{2}}
		& \cdots
		\\[6pt]
		\cdots & \cdots & \cdots 
		\\[6pt]
		\cdots & \cdots & 
		\dfrac{f_{\theta,\theta}}{g_\theta} - \dfrac{f_{\theta}\,g_{\theta,\theta}}{g_\theta^{2}} - 1  
	\end{bmatrix},
\end{equation}
where, for instance, \(f_{k,k} = \partial f_{k}/\partial k_{n}\), \(f_{k,\varepsilon} = \partial f_{k}/\partial \varepsilon_{n}\), and \(g_{\theta,\theta} = \partial g_{\theta}/\partial \overline{\theta^2}_{n}\).

Substituting \(k_{n}=f_{k}/g_{k}\), \(\varepsilon_{n}=f_{\varepsilon}/g_{\varepsilon}\), and \(\overline{\theta^2}_{n}=f_{\theta}/g_{\theta}\) at the equilibrium into \(J_{S}\) yields:
\begin{equation}
	J_{S} = 
	\begin{bmatrix}
		1/g_{k} & & \\[3pt]
		& 1/g_{\varepsilon} & \\[3pt]
		& & 1/g_{\theta}
	\end{bmatrix}
	\begin{bmatrix}
		f_{k,k} - k\,g_{k,k} - g_{k}
		& f_{k,\varepsilon} - k\,g_{k,\varepsilon} 
		& \cdots
		\\[6pt]
		\cdots & \cdots & \cdots 
		\\[6pt]
		\cdots & \cdots & 
		f_{\theta,\theta} - \overline{\theta^2}\,g_{\theta,\theta} - g_{\theta}
	\end{bmatrix}.
\end{equation}

Meanwhile, the equations for a transient simulation with a small time step can be written:
\begin{align}
	\dfrac{\Delta k}{\Delta t} = f_k - g_k\,k_n, \quad 
	\dfrac{\Delta \varepsilon}{\Delta t} = f_\varepsilon - g_\varepsilon  \varepsilon_{n}, \quad
	\dfrac{\Delta \overline{\theta^2}}{\Delta t} = f_\theta - g_\theta\,\overline{\theta^2}_n,
\end{align}
and the corresponding Jacobian matrix for the transient system \(J_{T}\) is:
\begin{equation}
	J_{T} =
	\begin{bmatrix}
		f_{k,k} - k\,g_{k,k} - g_{k}
		& f_{k,\varepsilon} - k\,g_{k,\varepsilon} 
		& \cdots
		\\[6pt]
		\cdots & \cdots & \cdots 
		\\[6pt]
		\cdots & \cdots & 
		f_{\theta,\theta} - \overline{\theta^2} g_{\theta,\theta} - g_{\theta}
	\end{bmatrix}.
\end{equation}
Thus, the relationship
\begin{equation}
	J_{S} = G\, J_{T},
\end{equation}
holds, where
\begin{equation}
	G 
	= 
	\begin{bmatrix}
		1/g_{k} & & \\
		& 1/g_{\varepsilon} & \\
		& & 1/g_{\theta}
	\end{bmatrix}.
\end{equation}

Because the model is already designed so that \(J_{T}\) meets the stable node criterion of Section \ref{sec:2.3} (all real parts of its eigenvalues are negative) and can be decomposed as
\begin{equation}
	J_{T} = Q^{-1} \Lambda Q,
\end{equation}
with all real parts of the diagonal matrix $\Lambda$ being negative, it follows that
\begin{equation}
	J_{S} = G\,J_{T} 
	= 
	Q^{-1} (G\,\Lambda) Q.
\end{equation}
Because all entries of the diagonal matrix \(G\) are positive (i.e., \(g_{k}, g_{\varepsilon}, g_{\theta} > 0\) by definition), all real parts of \(G\,\Lambda\) remain negative, ensuring stable convergence of the steady-state solver near the equilibrium.

For this reason, the present model is implemented so that source terms in the turbulence transport equations use values from step \((n)\), while sink terms are treated implicitly, multiplied by step \((n+1)\) values. Intuitively, for instance, one might consider the simple form \(k_{n+1} = (P_n/\varepsilon_n)\,k_{n}\): if \(P_{n} > \varepsilon_{n}\), then \(k_{n+1}\) should increase.

\subsection{Behavior of $f_{\varepsilon g}$ in vertical convection problems}

In the present study, \(f_{\varepsilon g}\) has been optimized for unstably stratified flows. To illustrate how the newly proposed \(f_{\varepsilon g}\) behaves in an exceptional scenario, natural convection problems in a vertically heated channel are also considered here.

First, the APE for vertical natural convection problems is to be calculated. Consider an enclosed, vertically heated domain of height \(L\) and width \(H\). Gravity acts in the \(-z\) direction (\(g_i = - g \,\delta_{i3}\)), and the fluid domain is defined by \(0 \le x \le H\) and \(0 \le z \le L\).
Isothermal boundary conditions are imposed as $T=0$ at $x=0$ and $T=\Delta$ at $x=H$.

In the pure conduction state, the fluid’s temperature field is
\begin{equation}
	T(x,z)= \Delta \dfrac{x}{H}.
\end{equation}
After adiabatic rearrangement, the rearranged temperature field \(T^{\ast}\) is
\begin{equation}
	T^{\ast}(x,z) = \Delta\,\dfrac{z}{L}.
\end{equation}
Hence, the available potential energy is
\begin{equation}
	E_a = \dfrac{g\,\beta}{L\,H} \int_{0}^{L} \int_{0}^{H} \left(T^{\ast} - T\right)\, z \,dx\,dz 
	= \dfrac{1}{12}\,g\,\beta\,\Delta\,L.
\end{equation}
For a vertical channel problem, letting \(L \to \infty\) yields \(E_a \to \infty\).

The newly introduced global function is written as
\begin{equation}
	f_{\varepsilon g} \;\sim\; \left(\dfrac{k}{E_a}\right)^{a}\;\cdots
\end{equation}
where \(a>0\), as constrained in Section~\ref{sec:2.3}. When \(E_a \to \infty\) in vertical channels, \(f_{\varepsilon g} \to 0\).

Much of the previous literature on buoyant RANS modeling \cite{zeman1976modeling,henkes1991natural,kenjerevs1995prediction,dol1999dns,peng1999computation,kenjerevs2005contribution,dehoux2017elliptic} has focused on vertical convection to optimize existing models. In vertical natural convection, gravity and the direction of turbulent heat transfer are largely aligned, so the buoyant production of turbulent kinetic energy—proportional to their inner product—remains relatively small. Consequently, turbulent kinetic energy is mainly produced by velocity gradients, meaning that even if the buoyant term in the \(\varepsilon\)-equation is set to zero, the simulation is barely affected.

This suggests that the newly proposed function \(f_{\varepsilon g}\) has little impact on RANS results for vertical convection but becomes more significant when modifying solutions for unstably stratified flows. As a result, the new model function \(f_{\varepsilon g}\) can be integrated seamlessly with existing RANS models that have already been optimized for vertical convection problems.

\section{Simulation Results}\label{sec:3}

In this section, two existing models are also tested for comparison:
\begin{enumerate}
	\item GDH \(k\)-\(\varepsilon\): The standard \(k\)-\(\varepsilon\) model by Launder and Sharma \cite{launder1974application}, combined with a gradient diffusion hypothesis (GDH) for the turbulent heat flux, as presented in Eq.~(\ref{eq:9}). The model constants are set to \(\Pran_t = 1.0\) and \(C_{\varepsilon g} = 1.44\).
	\item AFM \(k\)-\(\varepsilon\)-\(\overline{\theta^2}\): The \(k\)-\(\varepsilon\)-\(\overline{\theta^2}\) model with the algebraic flux model (AFM) proposed by Kenjereš and Hanjalić \cite{kenjerevs1995prediction}, using \(C_{\varepsilon g} = 1.44\). This model is identical to the present model when \(f_{\varepsilon g} = 1\) is applied.
\end{enumerate}
The relevant equations and other model constants can be found in Section~\ref{sec:2.1}.

The new model is identical to Case~2 but includes an additional global potential energy–coupled function \(f_{\varepsilon g}\), given in Eq.~(\ref{eq:16}) of Section~\ref{sec:2.2}. 
The constants of \(f_{\varepsilon g}\) are first optimized for turbulent Rayleigh–Bénard convection, and are set to \(C_{\varepsilon g} = 0.8\), \(a = 0.75\), \(b = 0.5\), \(d = 0.5\), and \(e = 0.4\). 
These constants are then verified in two types of internally heated problems.

\subsection{Numerical methods}

The current model is implemented in the open-source software OpenFOAM \cite{openfoam2025}. OpenFOAM employs the finite volume method on unstructured, collocated grids to solve fluid flow. Details on its numerical methods can be found in the publicly available source code \cite{openfoam2025} and in references such as \citet{moukalled2016finite}.

The pressure–velocity coupling is handled with the SIMPLE method. For discretization, Laplacian terms are treated using the “Gauss linear corrected” scheme, gradients with “Gauss linear,” and convection terms with the “bounded Gauss upwind” scheme. 
At no-slip walls, \(k\), \(\varepsilon\), and \(\theta\) are set to zero as Dirichlet boundary conditions, following the standard \(k\)-\(\varepsilon\) model by \citet{launder1974application}.

In all test cases, so-called steady-state simulation method is applied; all time derivative terms are neglected, and the model transport equations are iteratively solved as described in Section \ref{sec:2.3.4}. Moreover, validation across various tests has confirmed that even when a transient approach—where time derivatives are retained until convergence—is used, the same results are consistently obtained.

All problems are set up as one-dimensional; all variables depend only on \(z\) (parallel to gravity), with the \(x\) and \(y\) directions being periodic. Therefore, the mean velocity field is always zero in all simulations. In all one-dimensional simulation cases, even when the grid is refined to very small cells, consistent, grid-independent results are obtained. The grid resolution is set such that the height of the cells adjacent to the wall is less than approximately 1/100 of the thermal boundary layer thickness \(\delta_T\), and the cell-to-cell grid growth rate is kept below 1.05. Here, the thermal boundary layer thickness is estimated based on the measured Nusselt number using the heat balance relation \(\delta_T / L \approx \Nu^{-1}\). A more detailed description of the simulation setup can be found in the supplementary material.

Extensive testing under various scenarios confirms that the proposed model consistently predicts the same convergent values, even when initial conditions are altered. However, certain exceptions should be noted. For instance, if \(f_{\varepsilon g}\) is initially set to an extremely high value, \(\varepsilon\) may spike, forcing \(k\) to drop to zero and leading to an incorrect laminar prediction. To avoid such issues, it is important to ensure that \(f_{\varepsilon g}\) calculated from the initial conditions is not excessively large.
The following setup is recommended for the initial values of \(k\), \(\varepsilon\), and \(\overline{\theta^{2}}\): (i) Use a conduction-based temperature profile as the initial temperature field to prevent the available potential energy \(E_a\) from becoming zero.
(ii) Set the initial values of \(k\) and \(\varepsilon\) so that \(\Rey_t = k^{2}/(\nu \,\varepsilon)\) is in the range of approximately 10 to 100. This avoids very small \(\Rey_t\) values that would activate the damping functions in the standard \(k\)–\(\varepsilon\) model.
(iii) Choose initial values of \(k\), \(\varepsilon\), and \(\overline{\theta^{2}}\) so that the buoyant turbulent kinetic energy production is on the same order of magnitude as \(\varepsilon\) (or at least does not significantly exceed \(\varepsilon\)).

It is important to note that achieving grid independence in one-dimensional problems does not guarantee that the model will exhibit strict grid independence in two- or three-dimensional simulations. Many previous studies, such as those by \citet{peng1999computation,dol2001computational}, have reported that conventional RANS models are quite sensitive to grid resolution when applied to two- or three-dimensional natural convection problems. Furthermore, these models have been noted to be highly sensitive to initial conditions, making it difficult to obtain consistent results.

The reported sensitivity of the results to grid resolution and initial conditions can be illustrated by the following example: Suppose the grid is refined to a very fine resolution and the initial value for the turbulence model's \(\nu_t\) is set very low. During simulations, several thermal plumes may detach from the walls and become entangled with one another. Because basic models such as the standard \(k\)-\(\varepsilon\) model are known to perform very poorly in situations with significant mean streamline curvature \cite{wilcox1998turbulence,pope2000turbulent}, it is extremely difficult to guarantee the normal operation of the \(k\)-\(\varepsilon\) RANS model under such complex streamline conditions. In contrast, if the grid is set coarsely, small thermal plumes no longer appear, and only steady convection cells develop, which may lead to more stable results in the context of RANS calculations.

In summary, in higher-dimensional geometries, many intricate problems arise—not only in determining \(f_{\varepsilon g}\)—which further complicate the modeling. To eliminate all of these complicating factors, the present study conducts simulation validation in a one-dimensional setting. In one-dimensional problems, the severe issues related to initial conditions and grid dependency do not arise, and consequently, all simulation results presented here are almost independent of the initial conditions and grid resolution.

\subsection{Rayleigh–Bénard Convection}\label{sec:4.1}

\begin{figure}[h]
	\centering
	\begin{subfigure}{0.48\textwidth}
		\centering
		\includegraphics[scale=0.8]{"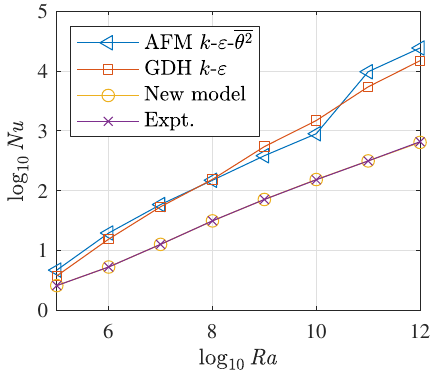"}
		\caption{All model results}
	\end{subfigure}
	\begin{subfigure}{0.48\textwidth}
		\centering
		\includegraphics[scale=0.8]{"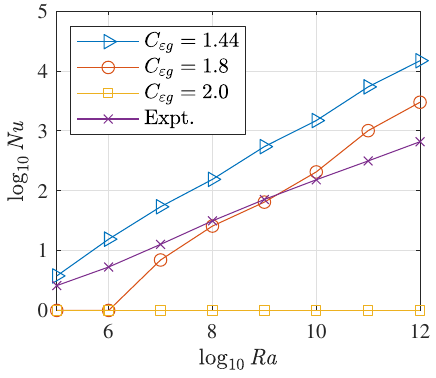"}
		\caption{GDH \(k\)–\(\varepsilon\), varying \(C_{\varepsilon g}\)}
	\end{subfigure}	
	\begin{subfigure}{0.48\textwidth}
		\centering
		\includegraphics[scale=0.8]{"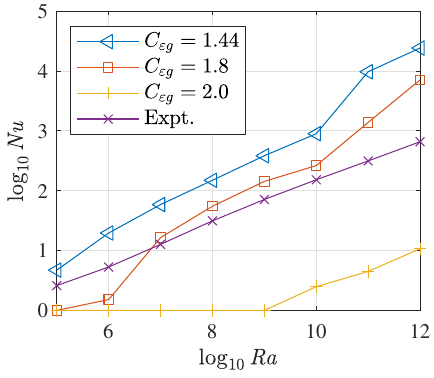"}
		\caption{AFM \(k\)–\(\varepsilon\)–\(\overline{\theta^2}\), varying \(C_{\varepsilon g}\)}
	\end{subfigure}
	\begin{subfigure}{0.48\textwidth}
		\centering
		\includegraphics[scale=0.8]{"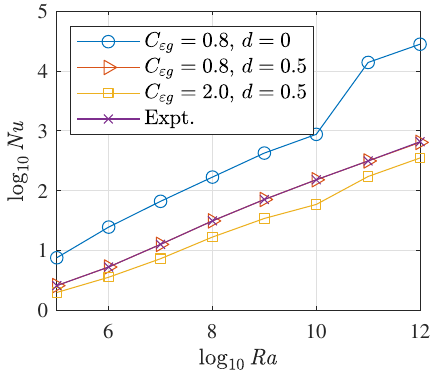"}
		\caption{New model, varying \(C_{\varepsilon g}\) and \(d\)}
	\end{subfigure}
	\caption{\(\Nu\) dependence on \(\Ra\) with fixed \(\Pran=1\) in Rayleigh–Bénard convection.}
	\label{fig:4}
\end{figure}

Figure~\ref{fig:4} illustrates the simulated dependence of \(\Nu\) on \(\Ra\) at a fixed \(\Pran = 1\) in Rayleigh–Bénard convection. The experimental data shown in the figures follow the correlation \(\Nu = 0.124\,\Ra^{0.309}\), as reported by Niemela et al.~\cite{niemela2000turbulent}.

In Figure~\ref{fig:4}(a), the new model \((C_{\varepsilon g} = 0.8,\, a = 0.75,\, \dots)\) is compared with two existing models \((C_{\varepsilon g} = 1.44)\). It is apparent that the new model reproduces the Rayleigh-number dependence of the Nusselt number over a wide range of parameters, whereas the other existing models do not.

Even if \(C_{\varepsilon g}\) is varied in the existing models, matching the experimental data is challenging. Figures~\ref{fig:4}(b) and \ref{fig:4}(c) show results for the two existing models while varying \(C_{\varepsilon g}\). Modifying \(C_{\varepsilon g}\) shifts the overall magnitude of \(\Nu\) across all \(\Ra\) values, but the scaling exponent \(n\) in \(\Nu \sim \Ra^n\) cannot be adjusted properly.

Figure~\ref{fig:4}(d) presents results for the new model, where \(d\) and \(C_{\varepsilon g}\) are varied while \(a=0.75\) is held fixed. In contrast to the existing models, the new global function \(f_{\varepsilon g}\) allows close agreement with the experimental data. As shown in Section~\ref{sec:2.2}, the simulated exponent \(n\) can be modified by varying \(a\) and \(d\). Changing \(C_{\varepsilon g}\) shifts the overall \(\Nu\) magnitude across the range of \(\Ra\). Adjusting \(b\) has almost no effect, because buoyant production and the dissipation rate balance each other.

\begin{figure}[h]
	\centering
	\begin{subfigure}{0.48\textwidth}
		\centering
		\includegraphics[scale=0.8]{"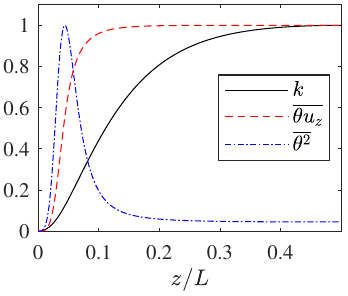"}
		\caption{}
	\end{subfigure}
	\begin{subfigure}{0.48\textwidth}
		\centering
		\includegraphics[scale=0.8]{"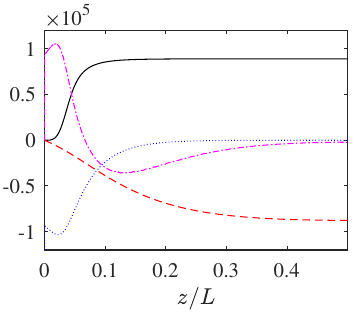"}
		\caption{}
	\end{subfigure}	
	\caption{The simulation results from the new model in Rayleigh–Bénard convection for \(\Ra=10^8\) and \(\Pran=10^{2}\):  
		(a) \(k\), \(\overline{\theta u_z}\), and \(\overline{\theta^2}\), normalized by their maximum values.  
		(b) The budget of the \(k\)-equation, nondimensionalized by \(L^4/\nu^3\): 
		Solid line (black), $g\beta \overline{\theta u_z}$;
		Dashed line (red), $-\varepsilon$;
		Dotted line (blue), $-\varepsilon_0$;
		Dash-dotted line (magenta), $\nabla \cdot (\nu+\nu_t / \sigma_k) \nabla k$}
	\label{fig:5}
\end{figure}

Figure \ref{fig:5} shows the simulation results for \(\Ra = 10^8\) and \(\Pran = 10^2\). In Figure \ref{fig:5}(a), the distributions of \(k\), \(\overline{\theta u_z}\), and \(\overline{\theta^2}\) are illustrated. Figure \ref{fig:5}(b) presents the budget of the \(k\)-equation.
Although the current model does not include specialized near-wall treatment for buoyancy effects, it demonstrates that once the Nusselt number is accurately predicted, the distributions and budgets of all turbulence variables are also determined precisely. When turbulent heat transfer is properly captured, the heat balance equation yields an accurate prediction of the thermal boundary layer thickness. Consequently, the ratio between the momentum and thermal boundary layer thicknesses is determined by the Prandtl number, and the near-wall distributions of \(k\), \(\varepsilon\), and \(\overline{\theta^2}\) are primarily governed by these boundary layer thicknesses.

\begin{figure}[h]
	\centering
	\begin{subfigure}{0.48\textwidth}
		\centering
		\includegraphics[scale=0.8]{"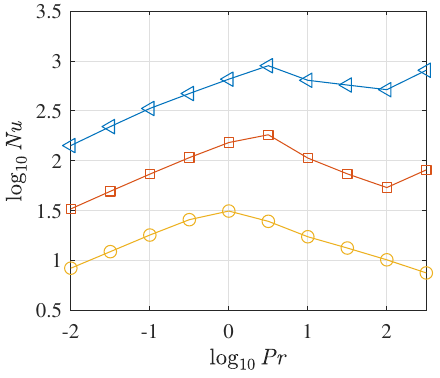"}
		\caption{New model ($\Ra=10^{8}, 10^{10}, 10^{12}$)}
	\end{subfigure}
	\begin{subfigure}{0.48\textwidth}
		\centering
		\includegraphics[scale=0.9]{"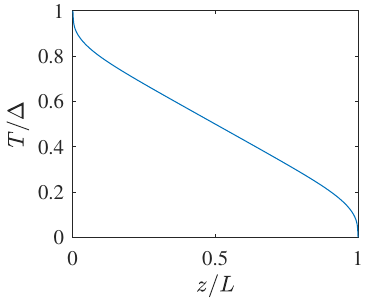"}
		\caption{$\Ra=10^{8}$, $\Pran=10^{-1}$}
	\end{subfigure}
	\begin{subfigure}{0.48\textwidth}
		\centering
		\includegraphics[scale=0.9]{"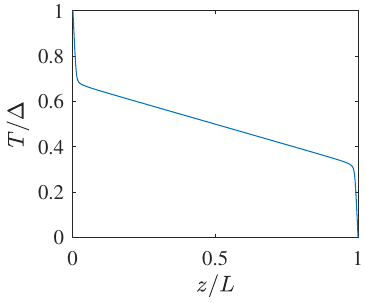"}
		\caption{$\Ra=10^{8}$, $\Pran=10^{0.5}$}
	\end{subfigure}	
	\begin{subfigure}{0.48\textwidth}
		\centering
		\includegraphics[scale=0.9]{"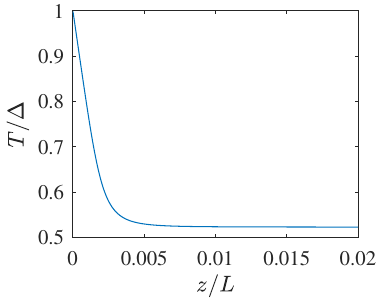"}
		\caption{$\Ra=10^{12}$, $\Pran=10^{2.5}$}
	\end{subfigure}
	\caption{Results from the new model: (a) Predicted dependence of \(\Nu\) on \(\Pran\) by the new model in Rayleigh–Bénard convection. (b)–(d) Temperature distributions along the height.}
	\label{fig:6}
\end{figure}

Figure~\ref{fig:6}(a) illustrates the predicted dependence of \(\Nu\) on \(\Pran\) by the new model in Rayleigh–Bénard convection. In the proposed function \(f_{\varepsilon g}\), the Prandtl-number dependence is primarily governed by \(e\). The chosen value of \(e = 0.4\) aims to replicate the experimentally observed \(\Nu \sim \Pran^{-1/8}\) scaling \cite{ahlers2009heat} for \(\Pran \lesssim 1\). Conversely, for \(\Pran \gtrsim 1\), experiments \cite{ahlers2009heat,silano2010numerical} within the range \(\Ra \leq 10^{10}\) report a scaling close to \(\Nu \sim \Pran^{0}\). In the present modeling results, the regime of \(\Ra = 10^{12}\) and \(\Pran \gtrsim 1\) partially reproduces this trend.

However, with \(e = 0.4\), an inaccurate trend emerges: \(\Nu\) decreases and then increases again as \(\Pran\) exceeds 1. Section~\ref{sec:2.3} introduced a rough scaling relation derived from the equilibrium analysis of the model, which does not account for this decrease-then-increase behavior in \(\Nu\).

To investigate the model’s prediction of \(\Nu\) with respect to \(\Pran\), Figure~\ref{fig:6}(b)–(d) presents the temperature distributions predicted by the model for various combinations of \(\Ra\) and \(\Pran\). The decrease in \(\Nu\) for \(\Pran \gtrsim 1\) is attributed to the linear combination of temperature gradient and buoyancy effects in the current implementation of the AFM, approximated as follows:
\begin{equation*}
	\overline{\theta u_z} \approx -\dfrac{2}{3} C_{\theta} \dfrac{k^2}{\varepsilon + \varepsilon_0} \dfrac{\partial T}{\partial z} + C_{\theta} \eta \beta g \dfrac{k}{\varepsilon + \varepsilon_0} \overline{\theta^2}
\end{equation*}
Depending on the dominant term on the right-hand side, the simulation is predicted to fall into one of three regimes:
\begin{enumerate}
	\item Low-\(\Pran\) regime (\(\Pran \lesssim 1\)): As shown in Figure~\ref{fig:6}(b), the modeled temperature gradient is nearly uniform at $\Delta/L$. Under these conditions, \(E_a \approx \tfrac{1}{6}\,g\,\beta\,\Delta\,L\), and 
	\begin{equation*}
		\left| \dfrac{2}{3} C_{\theta} \dfrac{k^2}{\varepsilon + \varepsilon_0} \dfrac{\partial T}{\partial z}\right| \gg \left| C_{\theta }\eta \beta g \dfrac{k}{\varepsilon + \varepsilon_0}\ \overline{\theta^2} \right|
	\end{equation*}
	
	\item Moderate-\(\Pran\) regime (\(1 \lesssim \Pran \lesssim 10^{2}\)): As shown in Figure~\ref{fig:6}(c), at moderate  \(\Pran\), the modeled temperature gradient in the center region diminishes as  \(\Pran\) increases. This reduces \(E_a\), leading to a decrease in \(\Nu\) with increasing  \(\Pran\). In this regime,
	\begin{equation*}
		\left| \dfrac{2}{3} C_{\theta} \dfrac{k^2}{\varepsilon + \varepsilon_0} \dfrac{\partial T}{\partial z} \right| \sim \left| C_{\theta }\eta \beta g \dfrac{k}{\varepsilon + \varepsilon_0}\ \overline{\theta^2} \right|.
	\end{equation*}
	
	\item High-\(\Pran\) regime (\(\Pran \gtrsim 10^{2}\)): As shown in Figure~\ref{fig:6}(d), at high  \(\Pran\), the temperature gradient is nearly zero except in the boundary-layer region near the walls. Under these conditions, $E_a \sim g\beta \Delta \delta_T$,	where \(\delta_T\) is the thermal boundary-layer thickness. Here, \(\Nu\) increases with increasing \(\Pran\), as estimated in Eq.~(\ref{eq:Equil 2}). In this regime,
	\begin{equation*}
		\left| \dfrac{2}{3} C_{\theta} \dfrac{k^2}{\varepsilon + \varepsilon_0} \dfrac{\partial T}{\partial z} \right| \ll \left| C_{\theta }\eta \beta g \dfrac{k}{\varepsilon + \varepsilon_0}\ \overline{\theta^2} \right|.
	\end{equation*}
\end{enumerate}

In summary, the reason the new model does not smoothly reproduce the \(\Nu \sim \Pran^{0}\) dependence for \(\Pran \gtrsim 1\) is related to the linear form of the reduced algebraic heat flux model currently employed. There remains potential for future improvements if the turbulent heat flux model is further refined.

\subsection{Internally Heated Convection}\label{sec:4.2}

\begin{figure}[h]
	\centering
	\begin{subfigure}{0.48\textwidth}
		\centering
		\includegraphics[scale=0.5]{"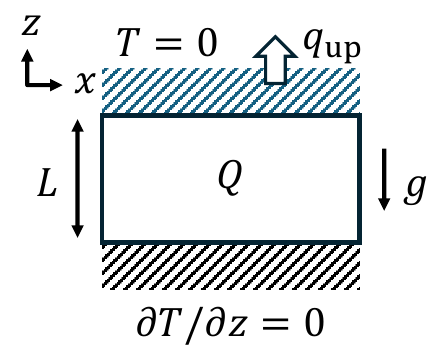"}
		\caption{Top cooling}
	\end{subfigure}
	\begin{subfigure}{0.48\textwidth}
		\centering
		\includegraphics[scale=0.5]{"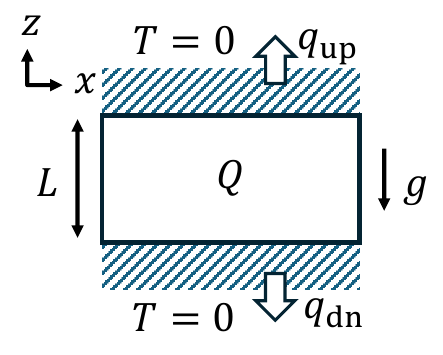"}
		\caption{Top and bottom cooling}
	\end{subfigure}
	\caption{The geometries and boundary conditions of two types of internally heated convection.}
	\label{fig:7}
\end{figure}

Using the optimized set of \(f_{\varepsilon g}\) model constants (\(C_{\varepsilon g} = 0.8\), \(a = 0.75\), \(b = 0.5\), \(d = 0.5\), and \(e = 0.4\)) from Section \ref{sec:4.1}, two types of internally heated convection problems are simulated. Similar to Rayleigh–Bénard convection, both cases feature zero net velocity across the domain, and turbulent kinetic energy is generated purely by buoyancy. Figure \ref{fig:7} illustrates the boundary conditions for these cases: (i) Top isothermal cooling with a bottom adiabatic no-slip wall. (ii) Top and bottom isothermal cooling with no-slip walls.

Here, \(Q\) \((\mathrm{K}/\mathrm{s})\) in Eq.~(\ref{eq:3}) denotes the rate of heat generation divided by the heat capacity per unit volume (i.e., the temperature rise per unit time caused by volumetric heating). The modified Rayleigh number \(\Ra^{\prime}\) and the Prandtl number \(\Pran\) are chosen as the dimensionless input parameters, following \citet{kulacki1977steady,goluskin2016internally}. The modified Rayleigh number is defined as
\begin{equation}
	\Ra^{\prime} = \dfrac{g\beta Q L^5}{\nu \alpha^2}.
\end{equation}

In these internally heated convection problems, all internally generated heat exits through the cooling surfaces. Because the total heat flux is fixed while the temperature difference is not known a priori, the Nusselt number cannot be defined as it is for Rayleigh–Bénard convection. Instead, previous studies \cite{kulacki1977steady,goluskin2016penetrative} have parameterized convective heat transfer using a dimensionless maximum temperature difference:
\begin{equation}
	T_{\max}^{\ast} = \dfrac{\alpha \,T_{\max}}{L^{2} \,Q}.
\end{equation}
The value of \(T_{\max}^{\ast}\) attains a maximum at \(\Ra^{\prime} = 0\) and subsequently decreases as \(\Ra^{\prime}\) increases. In the purely conductive state, \(T_{\max}^{\ast} = 0.5\) for case~(i) and \(T_{\max}^{\ast} = 0.125\) for case~(ii).  

For strategic comparison with other convective heat transfer problems, the Nusselt number can be defined as follows. Since \(T_{\max}^{\ast}\) is expressed as temperature divided by heat flux, \(\Nu\) scales with \((T_{\max}^{\ast})^{-1}\). If \(\Nu = 1\) is defined for the purely conductive state, then \(\Nu = 0.5\, (T_{\max}^{\ast})^{-1}\) for case (i) and \(\Nu = 0.125\, (T_{\max}^{\ast})^{-1}\) for case (ii). Although parameterizing the output in terms of the Nusselt number is often more intuitive—representing the ratio of conductive to convective heat transfer—this study presents results in terms of \(T_{\max}^{\ast}\), following the convention in previous works.

In case~(ii), the internally generated heat exits through both the top ($q_{up}$) and bottom ($q_{dn}$) cooling walls. Hence, the fraction of the bottom-wall heat flux relative to the total generated heat, denoted \(F_{dn}\), is introduced as an additional output parameter:
\begin{equation}
	F_{dn} = \dfrac{q_{dn}}{q_{up} + q_{dn}}.
\end{equation}

In the internally heated convection cases, the GDH \(k\)–\(\varepsilon\) model failed to produce properly converged results. Because internal heating and top cooling combine to create a region with a nearly zero temperature gradient yet substantial turbulent heat transfer, this scenario appears especially challenging for the GDH approach to simulate.

\subsubsection{Top cooling condition}

\begin{figure}[h]
	\centering
	\begin{subfigure}{0.48\textwidth}
		\centering
		\includegraphics[scale=0.8]{"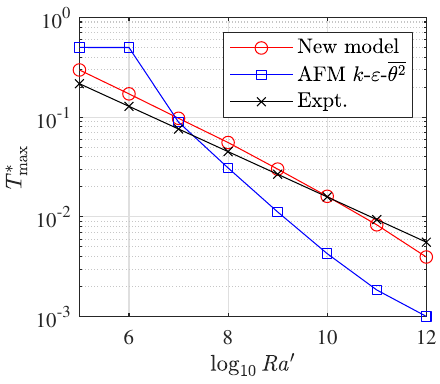"}
		\caption{}
	\end{subfigure}
	\begin{subfigure}{0.48\textwidth}
		\centering
		\includegraphics[scale=0.8]{"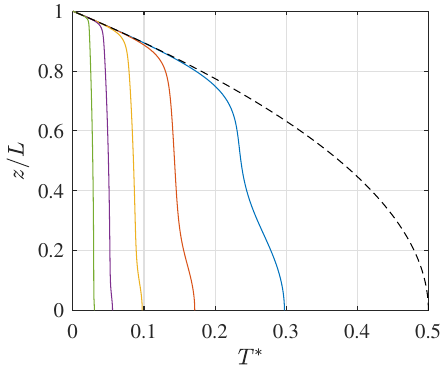"}
		\caption{}
	\end{subfigure}
	\caption{Internally heated convection with a top cooling condition at \(\Pran=6\) (water). (a) Predicted dependence of \(T_{\max}^{\ast}\) on \(\Ra^{\prime}\). (b) Dimensionless temperature profiles predicted by the new model along the vertical coordinate. Solid lines, from right to left, represent \(\Ra^{\prime}=10^{5}\), \(10^{6}\), \(10^{7}\), \(10^{8}\), and \(10^{9}\). The dashed line indicates the pure conduction state.}
	\label{fig:8}
\end{figure}

Figure~\ref{fig:8}(a) shows the RANS results for internally heated convection under a top cooling condition. The Prandtl number is set to \(\Pran=6\), following the experiments by Kulacki and Emara \cite{kulacki1977steady}, which are also plotted in the figure. Over the range \(\Ra^{\prime}=10^{5}\) to \(10^{12}\), the proposed model accurately predicts the dimensionless maximum temperature \(T_{\max}^{\ast}\). It is important to note that the model constants were originally optimized for Rayleigh–Bénard convection, without any specific adjustments for internal heating.

By contrast, the AFM \(k\)–\(\varepsilon\)–\(\overline{\theta^2}\) model significantly underpredicts \(T_{\max}^{\ast}\), showing a more rapid decrease of \(T_{\max}^{\ast}\) with increasing \(\Ra^{\prime}\) compared to the experiment. At \(\Ra^{\prime}=10^{12}\), this underprediction is about 55 times lower than the experimental value. This result is consistent with how the AFM \(k\)–\(\varepsilon\)–\(\overline{\theta^2}\) model overestimates the Nusselt number in Rayleigh–Bénard convection (see Figure~\ref{fig:6}a), given that \(T_{\max}^{\ast}\sim \Nu^{-1}\).

Figure~\ref{fig:8}(b) depicts the dimensionless temperature distribution \(\left(T^{\ast} = \alpha \,Q^{-1}\,L^{-2}\,T\right)\) for various \(\Ra^{\prime}\). As observed in experiments \cite{kulacki1977steady}, the temperature field remains nearly uniform except in the region close to the cooling wall.

\subsubsection{Top and bottom cooling condition}

\begin{figure}[h]
	\centering
	\begin{subfigure}{0.48\textwidth}
		\centering
		\includegraphics[scale=0.8]{"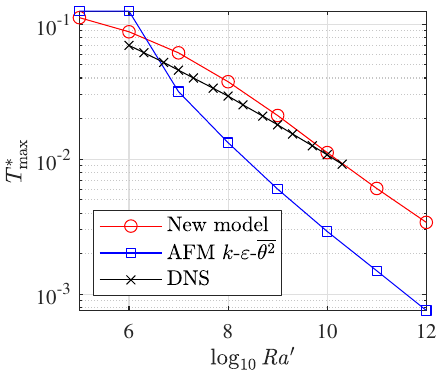"}
		\caption{}
	\end{subfigure}
	\begin{subfigure}{0.48\textwidth}
		\centering
		\includegraphics[scale=0.8]{"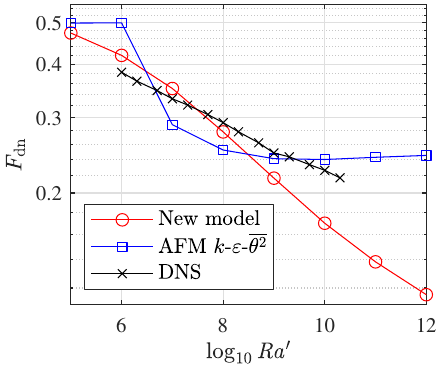"}
		\caption{}
	\end{subfigure}
	\caption{Internally heated convection with top and bottom cooling at fixed \(\Pran=1\).  
		(a) Predicted dependence of \(T_{\max}^{\ast}\) on \(\Ra^{\prime}\).  
		(b) Predicted dependence of \(F_{dn}\) on \(\Ra^{\prime}\).}
	\label{fig:9}
\end{figure}

\begin{figure}[h]
	\centering
	\begin{subfigure}{0.48\textwidth}
		\centering
		\includegraphics[scale=0.8]{"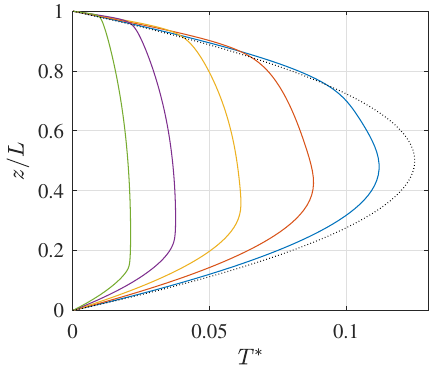"}
		\caption{}
	\end{subfigure}
	\begin{subfigure}{0.48\textwidth}
		\centering
		\includegraphics[scale=0.8]{"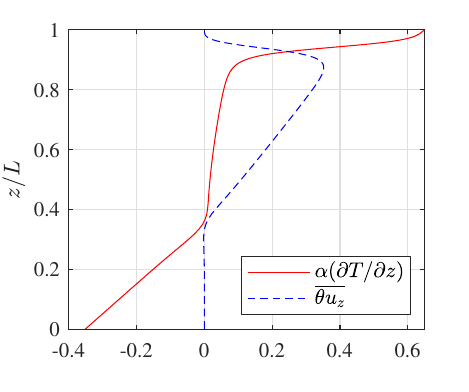"}
		\caption{}
	\end{subfigure}
	\caption{Results from the new model for internally heated convection with top and bottom cooling at fixed \(\Pran=1\). (a) Dimensionless temperature distribution predicted by the new model along the height. Solid lines (from right to left) represent \(\Ra^{\prime}=10^{5}\), \(10^{6}\), \(10^{7}\), \(10^{8}\), and \(10^{9}\). The dashed line indicates the pure conduction state.
		(b) Heat balance from the new model at \(\Ra^{\prime}=10^{7}\). All quantities are nondimensionalized by dividing by \(Q\,L\).}
	\label{fig:10}
\end{figure}

Figure~\ref{fig:9} presents RANS results for internally heated convection with top and bottom cooling at fixed \(\Pran=1\). The three-dimensional direct numerical simulation (DNS) data by Goluskin and van der Poel \cite{goluskin2016penetrative} are also shown for comparison. In Figures~\ref{fig:9}(a) and \ref{fig:9}(b), the new model predicts the dependence of \(T_{\max}^{\ast}\) and the downward heat fraction \(F_{dn}\) on \(\Ra^{\prime}\) more accurately than the existing AFM \(k\)–\(\varepsilon\)–\(\overline{\theta^2}\) model.

Although \(T_{\max}^{\ast}\) is well captured by the new model, the prediction of \(F_{dn}\) is less accurate, displaying a faster decrease as \(\Ra^{\prime}\) increases. This inaccuracy is believed to stem from the stably stratified region near the bottom of the domain, which is not explicitly modeled here but is simply clipped in the present approach. A log–log linear regression of the new model’s scaling for \(\Ra^{\prime}=10^{7}\)–\(10^{12}\) yields
\begin{equation*}
	T_{\max} \sim ({\Ra^{\prime}})^{-0.255} 
	\quad\text{and}\quad
	F_{dn} \sim ({\Ra^{\prime}})^{-0.0978}.
\end{equation*}
By contrast, the 3D DNS by Goluskin \textit{et al.} \cite{goluskin2016penetrative} indicates 
\begin{equation*}
	T_{\max} \sim ({\Ra^{\prime}})^{-0.205} 
	\quad\text{and}\quad
	F_{dn} \sim ({\Ra^{\prime}})^{-0.058}
	\quad\text{for}\quad 
	10^{6} \le \Ra^{\prime} \le 2\times10^{10}.
\end{equation*}

Figures~\ref{fig:10}(a) and \ref{fig:10}(b) show the dimensionless temperature \(T^{\ast}\) distribution and the heat balance predicted by the new model at \(\Ra^{\prime}=10^{7}\). Due to the cooling at the bottom wall, the flow becomes stably stratified for \(z/L\le 0.35\), as depicted in Figure~\ref{fig:10}(b).

To analyze the behavior of \(F_{dn}\) as predicted by the present model, assume that heat transfer in the stably stratified zone occurs primarily through conduction. If a control volume is defined below the zero-heat-flux point (e.g., \(z/L \le 0.35\) in Figure \ref{fig:10}b), all heat generated within this control volume exits exclusively through the bottom. Let \(h\) denote the height of this stably stratified control volume (e.g., \(h \approx 0.35\) in Figure \ref{fig:10}b). A heat balance then yields
\begin{equation*}
	Q (h - z) = \alpha \dfrac{\partial T}{\partial z},
\end{equation*}
and integrating \(dz\) from 0 to \(h\) and \(dT\) from 0 to \(T_{\max}\) yields
\begin{equation*}
	\dfrac{1}{2} Q h^{2}  = \alpha T_{\max}.
\end{equation*}
Hence,
\begin{equation}\label{eq:stable stratification estimate}
	F_{dn} = \dfrac{h}{L}  \sim  \left(T_{\max}^{\ast}\right)^{1/2}.
\end{equation}
This correlation aligns approximately with the new model’s predictions for \(F_{dn}\) and \(T_{\max}^{\ast}\).

By contrast, the 3D DNS results, \(T_{\max} \sim ({\Ra^{\prime}})^{-0.205}\) and \(F_{dn} \sim ({\Ra^{\prime}})^{-0.058}\), suggest that \(F_{dn}\) decreases more slowly in the DNS compared to the prediction of Eq.~(\ref{eq:stable stratification estimate}).
In reality, the stably stratified zone maintains a relatively small, but significant, turbulent heat transfer. Similar to Rayleigh–Bénard convection, a large-scale convection cell dominates the heat transfer across the entire domain, partly invading the stably stratified region. Thus, the large-scale circulation carries some hot fluid downward, producing a modest turbulent heat flux even in the stably stratified zone.
This mechanism explains the slow decrease of \(F_{dn}\) with increasing \(\Ra\) in 3D DNS, compared to its estimation based on the assumption of a fully conductive stably stratified zone.

Because modeling the stably stratified zone explicitly lies beyond the scope of this study, the present methodology sets negative buoyant production in the \(\varepsilon\)-equation to zero (a clipping approach). This choice likely drives the model to predict near-zero turbulent heat transfer in the stably stratified zone, resulting in the discrepancy in \(F_{dn}\). Properly accounting for stable stratification effects may improve the predictions.

\begin{figure}[h]
	\centering
	\begin{subfigure}{0.48\textwidth}
		\centering
		\includegraphics[scale=0.8]{"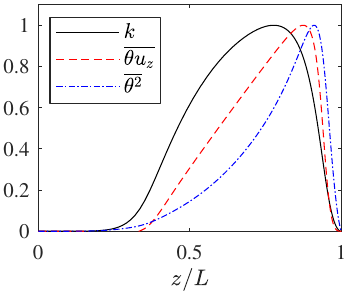"}
		\caption{}
	\end{subfigure}
	\begin{subfigure}{0.48\textwidth}
		\centering
		\includegraphics[scale=0.8]{"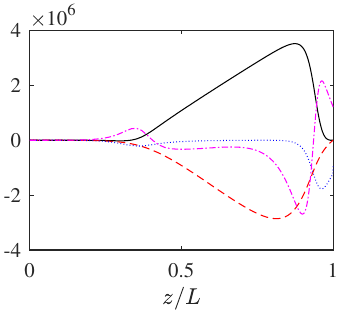"}
		\caption{}
	\end{subfigure}
	\caption{Results from the new model for internally heated convection with top and bottom cooling at \(\Ra^{\prime}=10^{7}\) and \(\Pran=1\). 
		(a) Profiles of \(k\), \(\overline{\theta u_z}\), and \(\overline{\theta^2}\), each normalized by its maximum value.  
		(b) Budget of the \(k\)-equation, nondimensionalized by \(\tfrac{L^4}{\nu^3}\): 
		black solid line, \(g\,\beta\,\overline{\theta u_z}\);
		red dashed line, \(-\varepsilon\);
		blue dotted line, \(-\varepsilon_0\);
		magenta dash-dotted line, \(\nabla \cdot \left[(\nu + {\nu_t}/{\sigma_k}) \nabla k\right]\).}
	\label{fig:11}
\end{figure}

Finally, Figure~\ref{fig:11} presents the profiles of the turbulent variables and the \(k\)-equation budget predicted by the new model. While the predicted budget aligns well with the 3D DNS data by Wörner et al.~\cite{worner1997direct} in the upper regions, the nearly zero values of \(k\), \(\varepsilon\), and \(\overline{\theta^2}\) in the stably stratified zone deviate from the DNS results. This discrepancy arises from the simplistic clipping approach applied to handle stable stratification in the current approach.

\section{Conclusion}\label{sec:4}

This study introduces a new buoyancy-correction approach for the \(\varepsilon\)-equation in \(k\)–\(\varepsilon\) RANS modeling by incorporating the concept of available potential energy (APE). A single algebraic function, \(f_{\varepsilon g}\), is added in front of the buoyancy-related term in the conventional \(\varepsilon\)-equation, thereby capturing global similarity relation of buoyant effects that standard one-point closures cannot address.

Because this new modeling concept relies on a global parameter, it requires a different approach than conventional one-point closures. The function \(f_{\varepsilon g}\) is assumed to be an arbitrary function of the independent turbulence variables, the material properties, and the global parameter APE. By applying a first-order Taylor expansion of \(f_{\varepsilon g}\) on a logarithmic scale, analytical scaling laws of \(\Nu \sim \Ra^{n}\,\Pran^{m}\) are derived through an equilibrium analysis based on a simplified Rayleigh–Bénard convection scenario. This procedure provides a basis for selecting new model constants so that the similarity relation observed in Rayleigh–Bénard convection can be reproduced.

Initial calibration and testing focus on Rayleigh–Bénard convection. Numerical simulations demonstrate that the model reproduces the Nusselt number across a wide range of Rayleigh and Prandtl numbers, offering significantly closer agreement with experimental data than the standard \(k\)–\(\varepsilon\) approach \cite{launder1974application} and \(k\)–\(\varepsilon\)–\(\overline{\theta^2}\) models employing algebraic turbulent heat flux formulations \cite{kenjerevs1995prediction}.

Further validation is carried out on two internally heated convection configurations. Conventional models exhibit pronounced discrepancies, notably underpredicting the maximum temperature. In contrast, the global similarity correction substantially reduces these errors, confirming the importance of incorporating global buoyant potential energy when nonlocal flow structures dominate heat transfer in natural convection flows.

The newly added function \(f_{\varepsilon g}\) vanishes in two situations: (i) when buoyancy is negligible, and (ii) when the APE per unit volume approaches infinity, as in a vertical channel domain. Consequently, in these exceptional cases, the modified model reverts entirely to the original \(k\)–\(\varepsilon\) formulation. As a result, the proposed function \(f_{\varepsilon g}\) can be seamlessly integrated with conventional RANS models.

It is important to note that the specific values for the model constants proposed for \(f_{\varepsilon g}\) and \(C_{\varepsilon g}\) in the current work are optimized for one-dimensional conditions using the model of \citet{kenjerevs1995prediction}. If \(f_{\varepsilon g}\) and \(C_{\varepsilon g}\) are combined with another model, their constants will likely need to be re-optimized. Nevertheless, due to its simplicity, the current modeling approach can be easily integrated into existing models based on \(k\)–\(\varepsilon\) or \(k\)–\(\omega\) RANS approaches, making it very practical for engineering applications. The core functionality of the proposed \(f_{\varepsilon g}\) is that it provides a new means of adjusting the power-law exponents in the \(\Nu\)–\(\Ra\)–\(\Pran\) correlation predicted by a given model. This is expected to greatly enhance both the performance and consistency of individual RANS models for buoyancy-driven turbulent flows.

\section*{Supplementary Material} 
The simulation source code is implemented as a solver application based on OpenFOAM v2046 (\url{https://github.com/DSJoo-CFD/RANS-model-for-buoyant-flows}). The supplementary material includes descriptions of the numerical schemes, simulation details, and data tables. 

\section*{ACKNOWLEDGMENTS}
This work did not receive any specific grant from funding agencies in the public, commercial, or not-for-profit sectors.

\section*{Conflict of Interest}
The author has no conflicts to disclose.

\section*{Data availability statement}
Data sharing is not applicable to this article as no new data were created or analyzed in this study.

\bibliography{NonlocalBuoyancyRANS}

\clearpage

\begin{center}
\textbf{\large Supplemental Materials}
\end{center}
\setcounter{equation}{0}
\setcounter{figure}{0}
\setcounter{table}{0}
\setcounter{page}{1}

\section*{I. Description for the Source Code}

This section outlines the numerical methods implemented for the newly added components. The current model builds upon (i) the standard \(k\)-\(\varepsilon\) model by Launder and Sharma \cite{launder1974application} and (ii) the reduced algebraic turbulent heat flux model (AFM) along with a temperature variance transport equation proposed by Kenjereš et al. \cite{kenjerevs1995prediction}. Although validation has so far been limited to one-dimensional cases, the code is designed to run in three dimensions as well.

OpenFOAM \cite{openfoam2025} already includes a RANS solver for buoyancy-driven flows, named “buoyantBoussinesqSimpleFoam.” This solver combines the standard \(k\)-\(\varepsilon\) model with the gradient diffusion hypothesis (GDH) for turbulent heat flux and incorporates buoyant turbulent kinetic energy production terms in both the \(k\)- and \(\varepsilon\)-equations. When integrating the new model, the portions shared with the “buoyantBoussinesqSimpleFoam” solver remain unchanged from the original OpenFOAM code. The new additions include the AFM, the \(\overline{\theta^2}\) transport equation, the calculation of available potential energy, and the proposed global similarity correction function.

The primary goal for model discretization is to ensure numerical stability. As is well known, more complex models—such as those incorporating algebraic turbulence closures like the AFM and additional transport equations (e.g., for \(\overline{\theta^2}\))—tend to be less stable than standard gradient diffusion–based models \cite{wilcox1998turbulence,durbin2011statistical}. Achieving numerical stability requires careful consideration of all potential interactions among the newly introduced model terms, their transport equations, and the boundary conditions. The remainder of this section identifies the sources of numerical instability encountered with the current model and presents various strategies to mitigate them.

\subsection{Discretization of the Algebraic Turbulent Heat Flux Model}

When the algebraic turbulent heat flux model is used in the temperature transport equation, the flux vector is split into a diffusion-like term and a remainder term to maintain consistency in the numerical scheme. The details are as follows:

First, the algebraic turbulent heat flux model is expressed as:
\begin{equation}
	\overline{\theta u_i^{\prime}} = -C_{\theta}\dfrac{k}{\varepsilon + \varepsilon_0}
	\left( 
	\overline{u^{\prime}_{i} u^{\prime}_{j}} \,\dfrac{\partial T}{\partial x_j}
	+ \xi\,\overline{\theta u^{\prime}_{j}} \,\dfrac{\partial U_i}{\partial x_j}
	+ \eta \,\beta\,g_i\,\overline{\theta^{2}}
	\right).
\end{equation}
By expressing the Reynolds stress in terms of \(\nu_t\) via an eddy viscosity model, the above can be rearranged as
\begin{equation}
	\overline{\theta u_j^{\prime}} 
	= -C_{\theta}\,\dfrac{k}{\varepsilon + \varepsilon_0}
	\left( 
	\delta_{ij} + C_{\theta}\,\dfrac{k}{\varepsilon}\,\xi\,\dfrac{\partial U_i}{\partial x_j}
	\right)^{-1}
	\left( 
	\left[ 
	\dfrac{2}{3}\,k \,\delta_{il} - 2\, \nu_{t}\, S_{il}
	\right]
	\dfrac{\partial T}{\partial x_l}
	+ \eta\,\beta\,g_i\,\overline{\theta^{2}}
	\right),
\end{equation}
where \(S_{ij} = \tfrac{1}{2}\left(\tfrac{\partial U_i}{\partial x_j} + \tfrac{\partial U_j}{\partial x_i}\right)\) is the strain-rate tensor.

To approximate the inverse matrix, the Neumann series is applied to the identity matrix \(I\) and a matrix \(A\):
\begin{equation}
	(I - A)^{-1} = I + A + A^{2} + \cdots \;\;\approx\;\; I + A.
\end{equation}
In this study, only the first-order approximation \((I + A)\) is used, which improves numerical stability under near-singular conditions where the determinant of the inverse matrix is close to zero. In homogeneous shear flow, \(\varepsilon/k\) and \(\lvert \partial U_i / \partial x_j \rvert\) exhibit proportional scaling under self-similarity. Thus, if the product of the model constant \(C_{\theta}\,\xi\) and the velocity gradient is sufficiently small, \(A\) remains small.

Using the first-order Neumann series approximation, the AFM becomes:
\begin{align}\label{eq:AFM expanded}
	\overline{\theta u_i} 
	=
	-
	\underset{\displaystyle{ \alpha_t \,\left(\partial T / \partial x_i\right) }}
	{\underbrace{
			\left( \dfrac{2}{3}\,C_{\theta} \,\dfrac{k^{2}}{\varepsilon+\varepsilon_0} \right)  \dfrac{\partial T}{\partial x_i}
	}}
	\;+\;
	\underset{\displaystyle{ \tilde{q_i} }}
	{\underbrace{
			\left[
			- C_{\theta}\,\eta\,\dfrac{k}{\varepsilon+\varepsilon_0}\,\beta\,g_i\,\overline{\theta^{2}} 
			\;+\; 2\,C_{\theta}\,\nu_{t}\,\dfrac{k}{\varepsilon+\varepsilon_0}\, S_{ij}\,\dfrac{\partial T}{\partial x_j }
			\;+\;\cdots
			\right]
	}},
\end{align}
where \(\alpha_t = (2/3)C_{\theta} k^{2} / (\varepsilon+\varepsilon_0) \).

The first term on the right, \(\alpha_t \nabla T\), is analogous to the GDH model. This term is combined with the molecular diffusivity \(\alpha\) to form \(\nabla \cdot [(\alpha + \alpha_t)\,\nabla T]\), evaluated by the Laplacian scheme.

The remaining term, \(\tilde{q_i}\), is handled separately. Because \(\tilde{q_i}\) contains temperature-gradient terms, computing \(\nabla \cdot \tilde{q_i}\) effectively requires taking the second derivative of the temperature field. If this is performed purely by central differencing on a colocated grid, checkerboard oscillations may appear in the temperature field due to odd–even decoupling. This phenomenon is analogous to the one encountered when solving the pressure Poisson equation on a colocated grid, which requires specialized treatments such as Rhie–Chow interpolation \cite{ferziger2002computational}.

To mitigate this issue, a simple interpolation method is adopted. The procedure involves interpolating gradients onto face centers to reduce odd–even decoupling, as follows:
(i) In OpenFOAM, colocated variables are stored at cell centers.
(ii) The Green–Gauss method is used to compute \(\nabla T\) at cell centers, and these gradients are substituted into the expression for \(\tilde{q_i}\) to obtain cell-centered values.
(iii) The cell-centered \(\tilde{q_i}\) is then interpolated to face centers.
(iv) The face-centered \(\tilde{q_i}\) is interpolated back to cell centers.
(v) The Green–Gauss method is used to compute \(\nabla \cdot \tilde{q_i}\) at cell centers for the temperature equation.
(vi) In the finite-volume temperature equation, the gradient \(\nabla \cdot \tilde{q_i}\) is evaluated by volume integration.

\subsection{Clipping Methods Based on Realizability to Prevent Spurious Oscillation}

\paragraph{Instability of temperature variance near an adiabatic wall.}

In the current model system, which solves the AFM and the \(\overline{\theta^2}\)-equation, there is a risk that \(\overline{\theta^2}\) may be spuriously amplified near the adiabatic wall. The cause of this numerical error is analyzed here, and a solution is proposed.

Along the wall-normal direction near an adiabatic wall, the total heat flux is 
\begin{equation}\label{eq:3.1.2-1}
	\overline{\theta u_z} - \alpha \dfrac{\partial T}{\partial z} \approx 0.
\end{equation}
The AFM is written as
\begin{equation}\label{eq:3.1.2-2}
	\overline{\theta u_z} \approx -\alpha_t \dfrac{\partial T}{\partial z} + C_{\theta} \eta \beta g \dfrac{k}{\varepsilon +\varepsilon_0} \overline{\theta^{2}}.
\end{equation}
Substituting this into the heat balance equation near an adiabatic wall gives:
\begin{equation}\label{eq:3.1.2-3}
	(\alpha + \alpha_t) \dfrac{\partial T}{\partial z}
	=  
	C_{\theta} \eta \beta g \dfrac{k}{\varepsilon +\varepsilon_0} \overline{\theta^{2}}.
\end{equation}
Hence, the temperature gradient \(\partial T / \partial z\) and the temperature variance \(\overline{\theta^2}\) become strongly coupled.

In a steady-state simulation, the temperature variance equation can be written as:
\begin{equation}\label{eq:3.1.2-4}
	0 = \left( -2  \overline{\theta u_z}  \dfrac{\partial T}{\partial z} \right)_{n}
	- \left( \dfrac{1}{R}  \dfrac{\varepsilon + \varepsilon_0}{k}\right)_{n}  \overline{\theta^2}_{n+1}.
\end{equation}
Here, the subscripts \(n\) and \((n+1)\) denote the iteration steps. By substituting Eqs.~\eqref{eq:3.1.2-1}–\eqref{eq:3.1.2-3} into Eq.~\eqref{eq:3.1.2-4} to eliminate the temperature gradient, the following is obtained:
\begin{equation}\label{eq:3.1.2-5}
	\overline{\theta^2}_{n+1} = 
	\left[ 
	- 2 \alpha R \left(\dfrac{C_\theta \eta \beta g}{(\alpha+\alpha_t)}\right)^2  \dfrac{k^3}{(\varepsilon+\varepsilon_0)^3} 
	\right]_{n} 
	\left(\overline{\theta^2}_{n}\right)^2 .
\end{equation}

Suppose all variables except \(\overline{\theta^2}\) remain nearly constant during the iterative process. Equation \eqref{eq:3.1.2-5} therefore behaves like a fixed-point iteration of the form \(x_{n+1} = f(x_n)\), for which a sufficient condition for convergence is \(\left|\partial f(x)/\partial x\right|<1\) \cite{burden2010numerical}.
Applying this concept indicates that the temperature variance equation is stable if
\begin{equation}\label{eq:3.1.2-6}
	\left|
	4 \alpha R \left(\dfrac{C_\theta \eta \beta g}{(\alpha+\alpha_t)}\right)^2  \dfrac{k^3}{(\varepsilon+\varepsilon_0)^3} \overline{\theta^2}
	\right|
	< 1.
\end{equation}
However, this condition is not always strictly satisfied. If even a single point violates it, spurious oscillations in \(\overline{\theta^2}\) can be exponentially amplified, leading to divergence and simulation failure.

To prevent spurious oscillation, a clipping method for the turbulent heat flux is proposed. The realizability condition for the turbulent heat flux is
\begin{equation}\label{eq:3.1.2-7}
	0\le \dfrac{\left|\overline{\theta u_i}\right|^2}{2 k \overline{\theta^2}} \le 1.
\end{equation}
If \(\overline{\theta^2}\) is dominant in the AFM, the model for turbulent heat flux can be approximated as
\begin{equation}\label{eq:3.1.2-8}
	\overline{\theta u_z} \approx C_\theta \eta \beta g \dfrac{k \overline{\theta^2}}{(\varepsilon + \varepsilon_0)}.
\end{equation}
Substituting this estimate for \(\overline{\theta u_z}\) into the realizability condition gives:
\begin{equation}\label{eq:3.1.2-9}
	\left(
	C_{\theta} \eta \beta g \dfrac{k \overline{\theta^2}}{(\varepsilon + \varepsilon_0)}
	\right)^2 \dfrac{1}{2 k \overline{\theta^2}} \le 1.
\end{equation}

The objective is to clip the model constant \(\eta\) so that this realizability condition is always satisfied. Rearranging Eq.~\eqref{eq:3.1.2-9} for \(\eta\) yields:
\begin{equation}\label{eq:3.1.2-10}
	\eta \;\le\; \dfrac{\sqrt{2} (\varepsilon + \varepsilon_0)}{C_{\theta} \beta g \sqrt{k \overline{\theta^2}}}.
\end{equation}
Hence, the following clipping function for \(\eta=0.6\) is proposed:
\begin{equation}
	\eta 
	= 
	\min \left( 0.6,\; \dfrac{0.2 (\varepsilon+\varepsilon_0)}{C_{\theta} \beta g \sqrt{k \overline{\theta^2}}} \right).
\end{equation}
Because only one term in AFM is considered here and the rest are not, an empirical clipping coefficient of 0.2 is introduced to allow a margin for convergence stability.

Substituting the clipped \(\eta\) into the stability condition in Eq.~\eqref{eq:3.1.2-6} gives:
\begin{equation}
	\left|
	4 \alpha R \left(\dfrac{C_\theta \eta \beta g}{(\alpha+\alpha_t)}\right)^2 \dfrac{k^3}{(\varepsilon+\varepsilon_0)^3} \overline{\theta^2}
	\right|
	= 
	0.16 \dfrac{R}{C_{\theta}}\;\dfrac{\alpha \alpha_t}{(\alpha + \alpha_t)} 
	< 
	1.
\end{equation}
Here, \(\alpha \alpha_t / (\alpha + \alpha_t)^2 \le 0.25\), \(R=0.75\), and \(C_\theta=0.15\). Therefore, the proposed clipping method guarantees the stability condition.

In most cases, the clipping function is not activated across the entire domain and may only be triggered in a small region near an adiabatic wall, so its overall impact on the modeling results is negligible.

\paragraph{Instability of velocity-gradient--AFM interaction.}

The instability caused by the interaction between the velocity gradient and the algebraic turbulent heat flux model, as well as the method applied to alleviate it, are described here. In AFM, even when the model constant related to the velocity gradient is \(\xi=0\), the velocity gradient appears in the modeled Reynolds stress tensor \(\overline{u_i u_j}\). In the Neumann series for AFM in Eq.~(\ref{eq:AFM expanded}), the following terms exhibit the strongest velocity-gradient effect:
\begin{equation}
	\overline{\theta u_i} = \cdots 
	+ C_{\theta} \dfrac{k}{\varepsilon+\varepsilon_0} 
	\left[
	2 \nu_{t} S_{ij} \dfrac{\partial T}{\partial x_j}
	\right]
	+ \cdots
\end{equation}

In a few simulation cases, numerical instability has been observed according to the following sequence:
\begin{enumerate}
	\item A velocity gradient abruptly forms in a small region, producing negative buoyant production of turbulent kinetic energy (\(- g_i \beta \,\overline{\theta u_i} < 0\)).
	\item This negative production reduces the local turbulent kinetic energy.
	\item Consequently, the turbulent viscosity \(\nu_t = C_\mu f_{\mu} k^{2}/\varepsilon\) also decreases in that region compared to neighboring cells.
	\item Where the total shear stress \(\tau\) is approximately constant across neighboring cells, momentum conservation is expressed as \(\tau = (\nu + \nu_t)\,\tfrac{\partial U}{\partial y}\). Thus, a reduced \(\nu_t\) in a localized cell leads to a larger velocity gradient in that cell.
	\item This mechanism forms a positive feedback loop, causing the spurious velocity gradient to grow at that point.
\end{enumerate}

To prevent this spurious feedback mechanism, the modeled Reynolds stress in the turbulent heat flux model is clipped based on a realizability condition. The Reynolds stress anisotropy \(b_{ij}\) is modeled as
\begin{equation}
	b_{ij} \equiv \dfrac{\overline{u^{\prime}_{i} u^{\prime}_{j}}}{2k} - \dfrac{1}{3}\,\delta_{ij} = \dfrac{\nu_{t}}{k} \,S_{ij}.
\end{equation}
As a rough realizability condition for \(b_{ij}\), its magnitude can be bounded as follows \cite{lumley1979computational}:
\begin{equation}
	0 \le b_{ij} b_{ji} \le \dfrac{2}{3}.
\end{equation}
Hence,
\begin{equation}
	0 \le \dfrac{\nu_t^2}{k^2} |S|^2 \le \dfrac{2}{3},
\end{equation}
where \(\left|S\right|^2 = S_{ij} S_{ji}\).

To satisfy the realizability condition, a clipping function is proposed:
\begin{equation}
	b_{ij} =  \dfrac{\nu_{t} \,k^{-1} S_{ij}}{\max\left(1,\, 2\,\nu_t \,k^{-1} |S|\right)}.
\end{equation}
Finally, the Reynolds stress in the AFM is modified to:
\begin{equation}
	\overline{\theta u_i} 
	= 
	-C_{\theta} \dfrac{k}{\varepsilon + \varepsilon_0}
	\left( 
	\delta_{ij} +  C_{\theta} \dfrac{k}{\varepsilon} \xi \dfrac{\partial U_i}{\partial x_j} 
	\right)  
	\left( 
	\left[ \dfrac{2}{3}\,k\,\delta_{jl} 
	- \dfrac{2\,\nu_t\,S_{jl}}{\max\left(1,\,2\,\nu_t\,k^{-1}|S|\right)}\right]
	\dfrac{\partial T}{\partial x_l}
	+ \eta \beta\,g_j\,\overline{\theta^2}
	\right) .
\end{equation}
This clipping method was applied only to AFM, not to the Reynolds stress in the momentum equation. Through this approach, the non-physical instability caused by velocity-gradient–AFM interaction is prevented.

\subsection{Calculation of Available Potential Energy}

\begin{figure}[h]
	\centering
	\includegraphics[scale=0.5]{"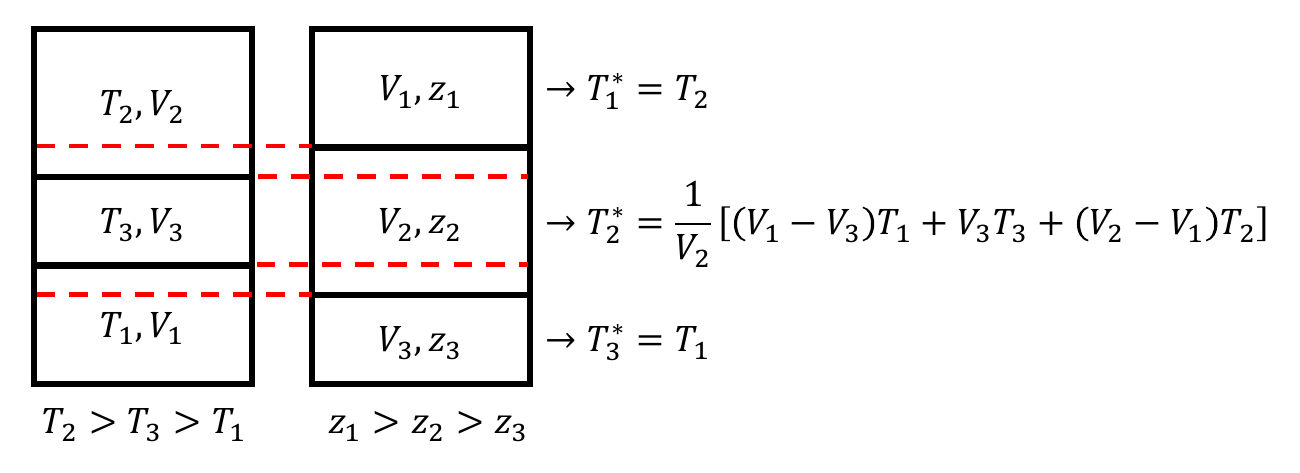"}
	\caption{Calculation of available potential energy}
	\label{fig:APE}
\end{figure}

Figure~\ref{fig:APE} illustrates the process for calculating the available potential energy. Let \(T_i\), \(V_i\), and \(z_i\) denote the temperature, volume, and height (location in the direction opposite to gravity) of each cell, respectively. All data are stored at the centers of the grid elements.

The adiabatic rearranged temperature field is computed as follows:
\begin{enumerate}
	\item Sort the cells by temperature (left side in the figure) and by height (right side in the figure). This step requires \(\mathcal{O}(N \log N)\) operations for \(N\) grid elements. The \texttt{std::stable\_sort} function from the C++ algorithm library is used for this sorting.
	\item Compare the two sets of cell sizes and reconstruct the rearranged temperature field ($T^{\ast}_i$), as shown in the diagram. This step requires \(\mathcal{O}(N)\) operations.
\end{enumerate}

In this manner, APE can be computed with an overall cost of \(\mathcal{O}(N \log N)\) per step, which is significantly less than the cost of performing matrix inversions for the transport equations.

\section*{II. Simulation Details and Data Tables}

The simulation details and data obtained from the new model are summarized in Tables 1 to 3.

The initial temperature field is set to the conduction-state distribution, and the initial values for \(k\), \(\varepsilon\), and \(\overline{\theta^2}\) are uniformly constant. Although the convergence results of the present RANS model are generally consistent regardless of the initial conditions, the values used are reported for the sake of objectivity. A steady-state simulation method is employed, with a minimum of \(10^5\) iterations performed. Data for each turbulence variable are extracted at every step, and the simulation is continued until convergence is clearly achieved.

The problems are defined as one-dimensional in the domain \(0 < z < L\), where \(z\) is the coordinate parallel to gravity and \(L\) is the reference length scale. The computational grids are designed so that from \(0 < z < L/2\), each maintains a constant growth rate, with the remainder of the domain constructed in mirror symmetry. Key parameters, such as the total number of grid elements, the width of the wall-attached cell (nondimensionalized by dividing by \(L\)), and the cell-to-cell expansion ratio, are documented.

\renewcommand{\arraystretch}{0.6}
\begin{table}[h]
	\caption{New model results in Rayleigh--B{\'e}nard convection}
	\makebox[\textwidth][c]{%
		\centering
		\begin{tabular}{|ccc|ccc|ccc|}
			\hline
			\multicolumn{3}{|c|}{Results}                           & \multicolumn{3}{c|}{Initial values}                  & \multicolumn{3}{c|}{Computational mesh}                                                                                                                                                                                                                   \\ \hline
			$\log_{10}\Ra$ 
			& $\log_{10}\Pran$ 
			& $\Nu$ 
			& $(L^2/\nu^2)k$ 
			& $(L^4/\nu^3)\varepsilon$ 
			& $\overline{\theta^2}/\Delta^2$ 
			& {\begin{tabular}[c]{@{}c@{}}Total\\ number\end{tabular}} 
			& {\begin{tabular}[c]{@{}c@{}}First\\ width\end{tabular}}
			& {\begin{tabular}[c]{@{}c@{}}Expansion\\ ratio\end{tabular}} \\
			5&0&$2.600$&$1.0\times10^{3}$&$3.2\times10^{5}$&$5.0\times10^{-2}$&200&$4.0\times10^{-4}$&1.040\\
			6&0&$5.313$&$1.0\times10^{6}$&$1.0\times10^{10}$&$1.0\times10^{-2}$&500&$9.3\times10^{-5}$&1.019\\
			7&0&$1.270\times10^{1}$&$1.0\times10^{7}$&$3.2\times10^{11}$&$1.0\times10^{-2}$&500&$9.3\times10^{-5}$&1.019\\
			8&0&$3.136\times10^{1}$&$1.0\times10^{6}$&$1.0\times10^{10}$&$1.0\times10^{-3}$&1000&$1.2\times10^{-5}$&1.012\\
			9&0&$7.140\times10^{1}$&$1.0\times10^{7}$&$3.2\times10^{11}$&$4.0\times10^{-4}$&1000&$1.2\times10^{-5}$&1.012\\
			10&0&$1.525\times10^{2}$&$4.0\times10^{7}$&$1.5\times10^{12}$&$4.0\times10^{-3}$&2000&$3.4\times10^{-6}$&1.007\\
			11&0&$3.147\times10^{2}$&$2.0\times10^{8}$&$3.2\times10^{13}$&$2.0\times10^{-3}$&2000&$3.4\times10^{-6}$&1.007\\
			12&0&$6.584\times10^{2}$&$1.0\times10^{8}$&$1.0\times10^{14}$&$1.0\times10^{-5}$&2000&$3.4\times10^{-6}$&1.007\\
			8&-2&$8.382$&$1.0\times10^{7}$&$1.0\times10^{12}$&$9.5\times10^{-8}$&1000&$1.2\times10^{-5}$&1.012\\
			8&-1.5&$1.228\times10^{1}$&$5.6\times10^{6}$&$3.2\times10^{11}$&$9.5\times10^{-8}$&1000&$1.2\times10^{-5}$&1.012\\
			8&-1&$1.802\times10^{1}$&$3.2\times10^{6}$&$1.0\times10^{11}$&$9.5\times10^{-8}$&1000&$1.2\times10^{-5}$&1.012\\
			8&-0.5&$2.569\times10^{1}$&$1.8\times10^{6}$&$3.2\times10^{10}$&$9.5\times10^{-8}$&1000&$1.2\times10^{-5}$&1.012\\
			8&0.5&$2.473\times10^{1}$&$5.6\times10^{5}$&$3.2\times10^{9}$&$9.5\times10^{-8}$&1000&$1.2\times10^{-5}$&1.012\\
			8&1&$1.732\times10^{1}$&$3.2\times10^{5}$&$1.0\times10^{9}$&$9.5\times10^{-8}$&1000&$1.2\times10^{-5}$&1.012\\
			8&1.5&$1.335\times10^{1}$&$1.8\times10^{5}$&$3.2\times10^{8}$&$9.5\times10^{-8}$&1000&$1.2\times10^{-5}$&1.012\\
			8&2&$1.017\times10^{1}$&$1.0\times10^{5}$&$1.0\times10^{8}$&$9.5\times10^{-8}$&1000&$1.2\times10^{-5}$&1.012\\
			8&2.5&7.510&$5.6\times10^{4}$&$3.2\times10^{7}$&$9.5\times10^{-8}$&1000&$1.2\times10^{-5}$&1.012\\
			10&-2&$3.295\times10^{1}$	&$1.0\times10^{8}$&$1.0\times10^{14}$&$9.5\times10^{-10}$&1000&$1.2\times10^{-5}$&1.012\\
			10&-1.5&$4.943\times10^{1}$&$5.6\times10^{7}$&$3.2\times10^{13}$&$9.5\times10^{-10}$&1000&$1.2\times10^{-5}$&1.012\\
			10&-1&$7.351\times10^{1}$	&$3.2\times10^{7}$&$1.0\times10^{13}$&$9.5\times10^{-10}$&1000&$1.2\times10^{-5}$&1.012\\
			10&-0.5&$1.076\times10^{2}$ &$1.8\times10^{7}$&$3.2\times10^{12}$&$9.5\times10^{-10}$&1000&$1.2\times10^{-5}$&1.012\\
			10&0.5&$1.824\times10^{2}$ &$5.6\times10^{6}$&$3.2\times10^{11}$&$9.5\times10^{-10}$&1000&$1.2\times10^{-5}$&1.012\\
			10&1&$1.066\times10^{2}$	&$3.2\times10^{6}$&$1.0\times10^{11}$&$9.5\times10^{-10}$&1000&$1.2\times10^{-5}$&1.012\\
			10&1.5&$7.398\times10^{1}$ &$1.8\times10^{6}$&$3.2\times10^{10}$&$9.5\times10^{-10}$&1000&$1.2\times10^{-5}$&1.012\\
			10&2&$5.392\times10^{1}$	&$1.0\times10^{6}$&$1.0\times10^{10}$&$9.5\times10^{-10}$&1000&$1.2\times10^{-5}$&1.012\\
			10&2.5&$8.106\times10^{1}$	&$5.6\times10^{5}$&$3.2\times10^{9}$&$9.5\times10^{-10}$&1000&$1.2\times10^{-5}$&1.012\\
			12&-2&$1.422\times10^{2}$	&$1.0\times10^{9}$	&$1.0\times10^{16}$	&$9.5\times10^{-12}$&1000&$1.2\times10^{-5}$&1.012\\
			12&-1.5&$2.203\times10^{2}$	&$5.6\times10^{8}$	&$3.2\times10^{15}$	&$9.5\times10^{-12}$&1000&$1.2\times10^{-5}$&1.012\\
			12&-1&$3.348\times10^{2}$	&$3.2\times10^{8}$	&$1.0\times10^{15}$	&$9.5\times10^{-12}$&1000&$1.2\times10^{-5}$&1.012\\
			12&-0.5&$4.726\times10^{2}$	&$1.8\times10^{8}$	&$3.2\times10^{14}$	&$9.5\times10^{-12}$&1000&$1.2\times10^{-5}$&1.012\\
			12&0.5&$8.985\times10^{2}$	&$5.6\times10^{7}$	&$3.2\times10^{13}$	&$9.5\times10^{-12}$&1000&$1.2\times10^{-5}$&1.012\\
			12&1&$6.417\times10^{2}$	&$3.2\times10^{7}$	&$1.0\times10^{13}$	&$9.5\times10^{-12}$&1000&$1.2\times10^{-5}$&1.012\\
			12&1.5&$5.751\times10^{2}$	&$1.8\times10^{7}$	&$3.2\times10^{12}$	&$9.5\times10^{-12}$&1000&$1.2\times10^{-5}$&1.012\\
			12&2&$5.167\times10^{2}$	&$1.0\times10^{7}$	&$1.0\times10^{12}$	&$9.5\times10^{-12}$&1000&$1.2\times10^{-5}$&1.012\\
			12&2.5&$8.089\times10^{2}$	&$5.6\times10^{6}$	&$3.2\times10^{11}$	&$9.5\times10^{-12}$&1000&$1.2\times10^{-5}$&1.012\\\hline
		\end{tabular}
	}
\end{table}

\renewcommand{\arraystretch}{0.9}
\begin{table}[h]
	\caption{New model results in internally heated convection with a top cooling wall, with \(\Pran=6\) fixed.}
	\makebox[\textwidth][c]{%
		\centering
		\begin{tabular}{|cc|ccc|ccc|}
			\hline
			\multicolumn{2}{|c|}{Results}                           & \multicolumn{3}{c|}{Initial values}                  & \multicolumn{3}{c|}{Computational mesh}                                                                                                                                                                                                                   \\ \hline
			$\log_{10}\Ra^{\prime}$ 
			& $T_{\max}^{\ast}$ 
			& $(L^2/\nu^2)k$ 
			& $(L^4/\nu^3)\varepsilon$ 
			& $\overline{\theta^2}/\Delta^2$ 
			& {\begin{tabular}[c]{@{}c@{}}Total\\ number\end{tabular}} 
			& {\begin{tabular}[c]{@{}c@{}}First\\ width\end{tabular}}
			& {\begin{tabular}[c]{@{}c@{}}Expansion\\ ratio\end{tabular}} \\
			5&$2.975\times10^{-1}$&$1.3\times10^{4}$&$1.7\times10^{6}$&$2.4\times10^{-5}$&200&$4.0\times10^{-4}$&1.040\\
			6&$1.714\times10^{-1}$&$4.1\times10^{4}$&$1.7\times10^{7}$&$3.8\times10^{-6}$&200&$2.3\times10^{-4}$&1.048\\
			7&$9.748\times10^{-2}$&$1.3\times10^{5}$&$1.7\times10^{8}$&$6.0\times10^{-7}$&200&$2.3\times10^{-4}$&1.048\\
			8&$5.572\times10^{-2}$&$4.1\times10^{5}$&$1.7\times10^{9}$&$9.5\times10^{-8}$&500&$5.3\times10^{-5}$&1.022\\
			9&$3.017\times10^{-2}$&$1.3\times10^{6}$&$1.7\times10^{10}$&$1.5\times10^{-9}$&500&$5.3\times10^{-5}$&1.022\\
			10&$1.610\times10^{-2}$&$4.1\times10^{6}$&$1.7\times10^{11}$&$2.4\times10^{-10}$&500&$5.3\times10^{-5}$&1.022\\
			11&$8.337\times10^{-3}$&$1.3\times10^{7}$&$1.7\times10^{12}$&$3.8\times10^{-11}$&500&$5.3\times10^{-5}$&1.022\\
			12&$3.964\times10^{-3}$&$4.1\times10^{7}$&$1.7\times10^{13}$&$6.0\times10^{-12}$&500&$5.3\times10^{-5}$&1.022\\ \hline
		\end{tabular}
	}
\end{table}

\begin{table}[h]
	\caption{New model results in internally heated convection with top and bottom cooling walls, with \(\Pran=1\) fixed.}
	\makebox[\textwidth][c]{%
		\centering
		\begin{tabular}{|ccc|ccc|ccc|}
			\hline
			\multicolumn{3}{|c|}{Results}                           & \multicolumn{3}{c|}{Initial values}                  & \multicolumn{3}{c|}{Computational mesh}                                                                                                                                                                                                                   \\ \hline
			$\log_{10}\Ra^{\prime}$ 
			& $T_{\max}^{\ast}$ 
			& $F_{dn}$ 
			& $(L^2/\nu^2)k$ 
			& $(L^4/\nu^3)\varepsilon$ 
			& $\overline{\theta^2}/\Delta^2$ 
			& {\begin{tabular}[c]{@{}c@{}}Total\\ number\end{tabular}} 
			& {\begin{tabular}[c]{@{}c@{}}First\\ width\end{tabular}}
			& {\begin{tabular}[c]{@{}c@{}}Expansion\\ ratio\end{tabular}} \\
			5&$1.120\times10^{-1}$&0.4729&$1.3\times10^{4}$&$1.7\times10^{6}$&$2.4\times10^{-5}$&200&$4.0\times10^{-4}$&1.040\\
			6&$8.799\times10^{-2}$&0.4193&$4.1\times10^{4}$&$1.7\times10^{7}$&$3.8\times10^{-6}$&200&$2.3\times10^{-4}$&1.048\\
			7&$6.144\times10^{-2}$&0.3508&$1.3\times10^{5}$&$1.7\times10^{8}$&$6.0\times10^{-7}$&200&$2.3\times10^{-4}$&1.048\\
			8&$3.755\times10^{-2}$&0.2778&$4.1\times10^{5}$&$1.7\times10^{9}$&$9.5\times10^{-8}$&500&$5.3\times10^{-5}$&1.022\\
			9&$2.110\times10^{-2}$&0.2166&$1.3\times10^{6}$&$1.7\times10^{10}$&$1.5\times10^{-9}$&500&$5.3\times10^{-5}$&1.022\\
			10&$1.121\times10^{-2}$&0.1700&$4.1\times10^{6}$&$1.7\times10^{11}$&$2.4\times10^{-10}$&500&$5.3\times10^{-5}$&1.022\\
			11&$6.081\times10^{-3}$&0.1381&$1.3\times10^{7}$&$1.7\times10^{12}$&$3.8\times10^{-11}$&500&$5.3\times10^{-5}$&1.022\\
			12&$6.410\times10^{-3}$&0.1158&$4.1\times10^{7}$&$1.7\times10^{13}$&$6.0\times10^{-12}$&500&$5.3\times10^{-5}$&1.022\\ \hline
		\end{tabular}
	}
\end{table}

\end{document}